\documentclass[
reprint,
superscriptaddress,
amsmath,amssymb,
aps,
pra,
]{revtex4-2}

\usepackage{graphicx}% Include figure files
\usepackage{dcolumn}% Align table columns on decimal point
\usepackage{bm}% bold math
\usepackage{hyperref}% add hypertext capabilities
\usepackage{graphicx} % Required for inserting images
\usepackage{xcolor}
\usepackage{makecell}
\usepackage{amsmath}
\usepackage{soul}
\DeclareMathOperator{\Tr}{Tr}
\DeclareMathOperator{\Det}{Det}
\mathchardef\mhyphen="2D

\usepackage{hyperref}

\newif\ifshowblue
\showbluefalse   % disable blue text

\newcommand{\bluemark}[1]{%
  \ifshowblue
    \textcolor{blue}{#1}%
  \else
    #1%
  \fi
}

\begin{document}

\title{Learning quantum tomography from incomplete measurements}

\author{Mateusz Krawczyk}
\affiliation{Institute of Theoretical Physics, 
%Faculty of Fundamental Problems of Technology, 
Wroc{\l}aw University of Science and Technology,
Wybrze\.{z}e Wyspia\'{n}skiego 27,
50-370 Wroc{\l}aw, Poland}

\author{Pavel Bal\'{a}\v{z}}
\affiliation{FZU - Institute of Physics of the Czech Academy of Sciences,
Na Slovance 1999/2, 182 00 Prague, Czech Republic}

\author{Katarzyna Roszak}
\affiliation{FZU - Institute of Physics of the Czech Academy of Sciences,
Na Slovance 1999/2, 182 00 Prague, Czech Republic}

\author{Jaros\l{}aw Paw\l{}owski}
\email{jaroslaw.pawlowski@pwr.edu.pl}
\affiliation{Institute of Theoretical Physics, 
%Faculty of Fundamental Problems of Technology,
Wroc{\l}aw University of Science and Technology,
Wybrze\.{z}e Wyspia\'{n}skiego 27,
50-370 Wroc{\l}aw, Poland}

\begin{abstract}
We revisit quantum tomography in an informationally incomplete scenario and propose improved state reconstruction methods using deep neural networks. In the first approach, the trained network predicts an optimal linear or quadratic reconstructor with coefficients depending only on the collection of (already taken) measurement operators. This effectively refines the undercomplete tomographic reconstructor based on pseudoinverse operation. 
The second, based on an LSTM recurrent network performs state reconstruction sequentially. 
It can also optimize the measurement sequence, which suggests a no-free-lunch theorem for tomography: by narrowing the state space, we gain the possibility of more efficient tomography by learning the optimal sequence of measurements. 
Numerical experiments for a 2-qubit system show that both methods outperform standard maximum likelihood estimation and also scale to larger 3- and 4-qubit systems.
Our results demonstrate that neural networks can effectively learn the underlying geometry of multi-qubit states and use it for their reconstruction.
%We revisit quantum tomography in an informationally incomplete measurement regime and propose improved state reconstruction methods based on deep neural networks. In the first approach, the trained network predicts an optimal linear or quadratic reconstructor whose coefficients depend only on the set of measurement operators already performed, thereby refining the standard pseudoinverse-based estimator for undercomplete tomography. The second approach employs a recurrent LSTM network that performs state reconstruction sequentially and can additionally optimize the measurement order. This adaptive behavior highlights a tradeoff between generality and efficiency in tomography: by exploiting prior structure in the state space, one can achieve more efficient reconstruction through learned measurement sequences. Numerical experiments for two-qubit systems show that both methods outperform standard maximum-likelihood estimation, with consistent performance gains observed also for three- and four-qubit systems. Our results demonstrate that neural networks can effectively learn geometric features of multi-qubit state spaces and leverage them to improve reconstruction from incomplete data.
\end{abstract}

\maketitle

\section{Introduction}
Quantum state tomography (QST)~\cite{Vogel1989,Kwiat2001,Paris2004} aims to determine the full state of a quantum system via a series
of quantum measurements.
Although the methods for complete state reconstruction are well known~\cite{Kwiat2001,Altepeter2005},
QST for incomplete data does not have a simple straightforward solution as the positivity criterion imposed on the state estimators makes analytical studies of such problems extremely difficult~\cite{Hradil2013}.
Approximate statistical-based methods for QST with an undercomplete or noisy measurement set are the maximum entropy (MaxEnt) principle~\cite{Buzek2004,Teo2011,Hradil2013},
maximum likelihood estimation (MLE)~\cite{Hradil1997,Hradil2001,Teo2011,Hradil2013},
or least-squares inversion~\cite{Opatrny1997}.

Although QST scales exponentially in the system size, approximate methods that scale much better (polynomially) already appear using either the singular value thresholding method~\cite{Cramer2010}, or neural networks (NNs)~\cite{Carrasquilla2019}.
In the case of permutationally invariant states (which occur in many practical situations, such as Dicke states or spin-squeezed states), exact tomography methods exist~\cite{Giza2010} that scale quadratically with system size~\cite{Giza2014}.
On the other hand, classical shadow tomography~\cite{Acharya2021,Huang2020} utilizes the fact that predicting important properties of a many-body state (without necessarily fully characterizing the quantum state) requires only a polynomially-growing number of measurements~\cite{Aaronson2020}.
Restricted Boltzmann machine (RBM)~\cite{Torlai2018,Melkani2020}, multilayer perceptron (MLP)~\cite{Xin2019,Kotuny2022}, convolutional NN~\cite{Schmale2022}, recurrent NN~\cite{Carrasquilla2019,Quek2021}, or conditional generative adversarial NN (CGAN)~\cite{Shahnawaz2021} fed by subsequent measurement snapshots that form an \textit{informationally complete} system can be used to reconstruct certain classes of states, but most models (beside that of Ref.~\cite{Quek2021}, or Ref.~\cite{Li2017}, but using integer programming) do not optimize the sequence of measurements. NN-aided tomography can also converge much faster~\cite{Kotuny2022,Shahnawaz2021} than other QST techniques, while attention-based NNs can be used to effectively denoise measurement data during QST~\cite{Palmieri2024}.
Deep NNs have also shown their usability in recognizing quantum entanglement~\cite{Pawlowski2024,Taghadomi2024,Chen2022,Asif2023,Urena2024}, or discord~\cite{Krawczyk2024}.

In this work, we focus on a different scenario: we examine various approaches to QST of general mixed states using \textit{undercomplete} measurement data, under the assumption that a single measurement expectation value is known exactly, and show that NNs outperform standard estimators such as MLE on a diverse test set ($S_\mathrm{test}$) of 2- to 4-qubit random mixed states.
We explicitly demonstrate that neural estimators yield optimal reconstruction for the 1-qubit case and identify the best measurement sequence for 2-qubit X states.

%\section{Quantum tomography}
\bluemark{In general, the expectation value of a projection-valued measurement (PVM) is given by the Born rule~\cite{Nielsen_Chuang_2010}
$m = \Tr(\Pi\rho)$,
where $\Pi$ is the projection operator and $\rho$ is the density matrix. Since $\Pi$ is a projector, this expectation value is also the probability of obtaining the outcome associated with $\Pi$. Experimentally, it is estimated from a finite number $N_c$ of copies as $\hat m = \frac{n_c}{N_c}$, where $n_c$ is a number of coincidence counts in a measurement setup.} For $N$ qubits, \bluemark{the expectation value} of joint measurements required for QST can be expressed as
\begin{equation}
    m_{i_1,..,i_N} = \!\Tr\!\left(\left(\mu_{i_1}\!\otimes\!\dots\!\otimes\mu_{i_N}\!\right)\rho\right)=\!\Tr\!\left(\Pi_{i_1,..,i_N}\rho\right),
    \label{eq:measurements}
\end{equation}
where $\mu_{i_k}=|\psi_{i_k}\rangle\langle \psi_{i_k}|$ %with $i_k \in \{0, 1, 2, 3\}$%
are single-qubit projectors acting on the $k$-th qubit state.
\bluemark{In the special case of rank-one projectors considered here, it can equivalently be interpreted as the corresponding measurement probability.}
%One can also rephrase the Kronecker string-product as $\mu_{i_1} \otimes \mu_{i_2} \otimes \dots \otimes \mu_{i_N}=\prod_{k=1}^{N}\mu_{i_k,k}$, where $\mu_{i_k,k}$ has nonidentity projector $|i_k\rangle\langle i_k|$ at $k$-th position only.
In the following, 
\bluemark{unless stated otherwise, we use the standard set of operators employed in photonic tomography by James \emph{et al.}~\cite{Kwiat2001} as a fixed catalog of single-qubit rank-one projectors,
\begin{equation}
|\psi_i\rangle \in
\left\{
|0\rangle,\ |1\rangle,\ 
\frac{|0\rangle+|1\rangle}{\sqrt{2}},\ 
\frac{|0\rangle-i|1\rangle}{\sqrt{2}}
\right\}.
\label{eq:single_qubit_catalogue}
\end{equation}
For an $N$-qubit system, this defines the product-projector catalog
\begin{equation}
\begin{aligned}
{\cal C}_N =
\Big\{
\Pi_\nu \equiv \Pi_{i_1,\ldots,i_N}
&=
\bigotimes_{k=1}^{N}
|\psi_{i_k}\rangle\langle\psi_{i_k}|:
i_k \in \{0,1,2,3\}
\Big\},
\end{aligned}
\label{eq:projector_catalogue}
\end{equation}
$|{\cal C}_N|=4^N$. The index $\nu$ is therefore a shorthand for the multi-index $(i_1,\ldots,i_N)$.
}

\bluemark{It is important to stress that the set in Eq.~\eqref{eq:single_qubit_catalogue} is not used as a single four-outcome PVM. In particular, the projectors onto $|0\rangle$ and $|1\rangle$ are complementary outcomes of a computational-basis measurement, and their probabilities are not independent once the trace of $\rho$ is fixed. Nevertheless, we keep both projectors in the catalog for two practical reasons. First, the resulting product catalog is the conventional James \emph{et al.} tomographic frame and leads to a simple linear reconstruction matrix in the fully measured case. Second, in the present work, the central object is not to use a complete set of PVMs but an incomplete list of expectation values $m_\nu=\mathrm{Tr}(\Pi_\nu\rho)$ chosen from a fixed, experimentally transparent catalog. Thus, the apparent redundancy of some probabilities is intentionally retained in the initial setting. When measurements are selected adaptively, the \emph{selector} model can learn to avoid redundant choices.
}

\bluemark{The projectors in Eq.~\eqref{eq:projector_catalogue} are Hermitian, idempotent, and positive semidefinite, but they do not form a PVM as they are not trace-orthonormal~\cite{yuan2022}. 
This does not prevent them from forming an informationally complete tomographic frame (the completeness of PVM is not the same as the completeness of QST itself).
Indeed, the Gramian matrix
\begin{equation}
G_{\nu\nu'}=\mathrm{Tr}(\Pi_\nu\Pi_{\nu'})
\end{equation}
is nonsingular, equivalently
\begin{equation}
\det G =
\det\!\left[\mathrm{Tr}(\mu_i\mu_j)\right]^{N4^{N-1}}
=
\left(\frac{1}{4}\right)^{N4^{N-1}}
\neq 0 ,
\end{equation}
hence operators $\Pi_\nu$ form a linearly independent set of dimension $4^N$  and give a complete measurement basis~\cite{Hradil2013}.
%Other informationally complete frames, for example, those based on Pauli settings or mutually unbiased bases, could be used instead. The James \emph{et al.} catalog is chosen here as a concrete and standard reference frame. The neural reconstruction methods studied below are mostly tied to this particular choice, except for the ``adjusted-basis'' selector model, which is allowed to propose an arbitrary projector.
}

After collecting \bluemark{a set of expectation values} $m_\nu$ one can tomographically reconstruct the state via the $\Gamma_\mu$-matrix basis~\cite{Kwiat2001} \bluemark{(Einstein summation convention assumed)}: 
\begin{equation}
\rho=\Gamma_\mu\left(\langle\psi_\nu|\Gamma_\mu|\psi_\nu\rangle\right)^{-1}_{\mu\nu}m_\nu,
\label{eq:fulltomo}
\end{equation}
where the tomographic state $|\psi_\nu\rangle=|\psi_{i_1}\rangle\otimes\!\dots\!\otimes|\psi_{i_N}\rangle$ is composed of states defining subsequent projectors, and $\Gamma_\mu=\bigotimes_{k=1}^N\sigma_{j_k}$ is the Kronecker product composed out of the Pauli matrices $\sigma_{j_k}$, with $\mu\equiv j_1,..,j_N$ and $j_k=0,1,2,3$ indexing the Pauli matrix of qubit $k$.
\bluemark{Note that $\sigma_0$, being the identity matrix, is included to complete the Pauli matrices.
Wherever possible, we adopt the Einstein summation convention. For instance, Eqs.~\ref{eq:fulltomo} and~\ref{eq:pseudoinverse} involves two contractions: with respect to the indices $\nu$ and $\mu$.}
The matrix $B_{\nu\mu}=\langle\psi_\nu|\Gamma_\mu|\psi_\nu\rangle$ relates subsequent measurement expectations $m_\nu$ with the state representation in the $\Gamma_\mu$ basis.
The choice of measurement basis ($\mu_{i_k}=|\psi_{i_k}\rangle\langle \psi_{i_k}|$) required to obtain full information about a quantum state $\rho$ is not unique, but 
%to meet the completeness requirement for the set of tomographic states $|\psi_\nu\rangle$~\cite{Kwiat2001}, 
$B_{\nu\mu}$ must be invertible.
$B_{\nu\mu}=\Tr((\bigotimes_{k=1}^N\mu_{i_k})(\bigotimes_{k=1}^N\sigma_{i_k}))=\bigotimes_{k=1}^N\Tr(\mu_{i_k}\sigma_{j_k})=\Tr(\mu_i\sigma_j)^{\otimes N}$, and $\Det(B_{\nu\mu})=\Det(\Tr(\mu_i\sigma_j))^{N4^{N-1}}$ is nonzero because $\Det(\Tr(\mu_i\sigma_j))=-2\neq 0$ in our case, so the condition is met.
\bluemark{Its inverse is denoted by $(B_{\nu\mu})^{-1}=B^{-1}_{\mu\nu}$ (note the index swap).}

QST is extremely resource-intensive, so it is natural to ask what happens if we do not have a complete set of measurements or if we have measurements with a significant uncertainty. \bluemark{Here we study the undercomplete regime in which fewer than a complete set of tomographic measurement expectation values is available~\cite{Hradil2013}. Operationally, this means that only $M$ values $m_{\nu}=\mathrm{Tr}(\Pi_{\nu}\rho)$ are provided to the reconstruction algorithm, with $M<4^N$, or if trace normalization is accounted for, fewer than the $4^N-1$ independent parameters required for a generic $N$-qubit density matrix. In the numerical results presented in this work, these values are evaluated exactly from the sampled target state $\rho$. Thus, the reported benchmarks correspond to the infinite-shot limit and isolate the error caused by informational incompleteness and by the reconstruction map itself. In an experiment, the same algorithms would take as input empirical estimates $\hat m_{\nu}=n_{\nu}/N_c$ instead of the exact expectation values $m_{\nu}$.}

When the number of measurement operators $M$ is smaller than the dimension of the complete set spanned by the $4^N$ $\Gamma_\mu$ matrices in the linear QST, $B_{\nu\mu}$ with dimension $M\times 4^N$ is no longer a square matrix, and becomes non-invertible.
The most natural way to keep the formula of Eq.~(\ref{eq:fulltomo}) operational is to use the pseudoinverse (Moore–Penrose inverse)~\cite{Penrose1955} operation: $(.)^+$. Since measurement expectation values are linearly independent, so will the rows of matrix $B_{\nu\mu}$ be, hence the formula for pseudoinverse: \bluemark{$B^+=B^\dag(BB^\dag)^{-1}$.}
Then the reconstruction formula is \bluemark{(Einstein summation assumed)}:
\begin{equation}
\rho\simeq\Gamma_\mu B^{+}_{\mu\nu}m_\nu.
\label{eq:pseudoinverse}
\end{equation}
Using the pseudoinverse as a QST estimator was also discussed in Ref.~[\onlinecite{Kotuny2022}].
In the results section we show that pseudoinverse-based reconstruction does not give the best possible solution, therefore, we aim to construct more accurate reconstruction schemes based on NN regressors.

\bluemark{It is important to emphasize that Eq.~(\ref{eq:pseudoinverse}) defines a raw linear reconstruction map. In the informationally incomplete regime this map is not constrained to return a physical density matrix: the reconstructed operator may fail to have unit trace and may also fail to be positive semidefinite. In the present work we do not post-process such reconstructions by trace normalization or eigenvalue clipping.
%or projection onto the set of physical density matrices. 
This choice is deliberate: our goal is to compare the reconstruction maps as estimators of the density operator, without adding an external correction step that could artificially improve the performance of a given method. %Consequently, deviations from normalization or positivity are counted as reconstruction errors rather than removed by post-processing.
}

\begin{figure}[t]
    \centering
    \includegraphics[width=.95\linewidth]{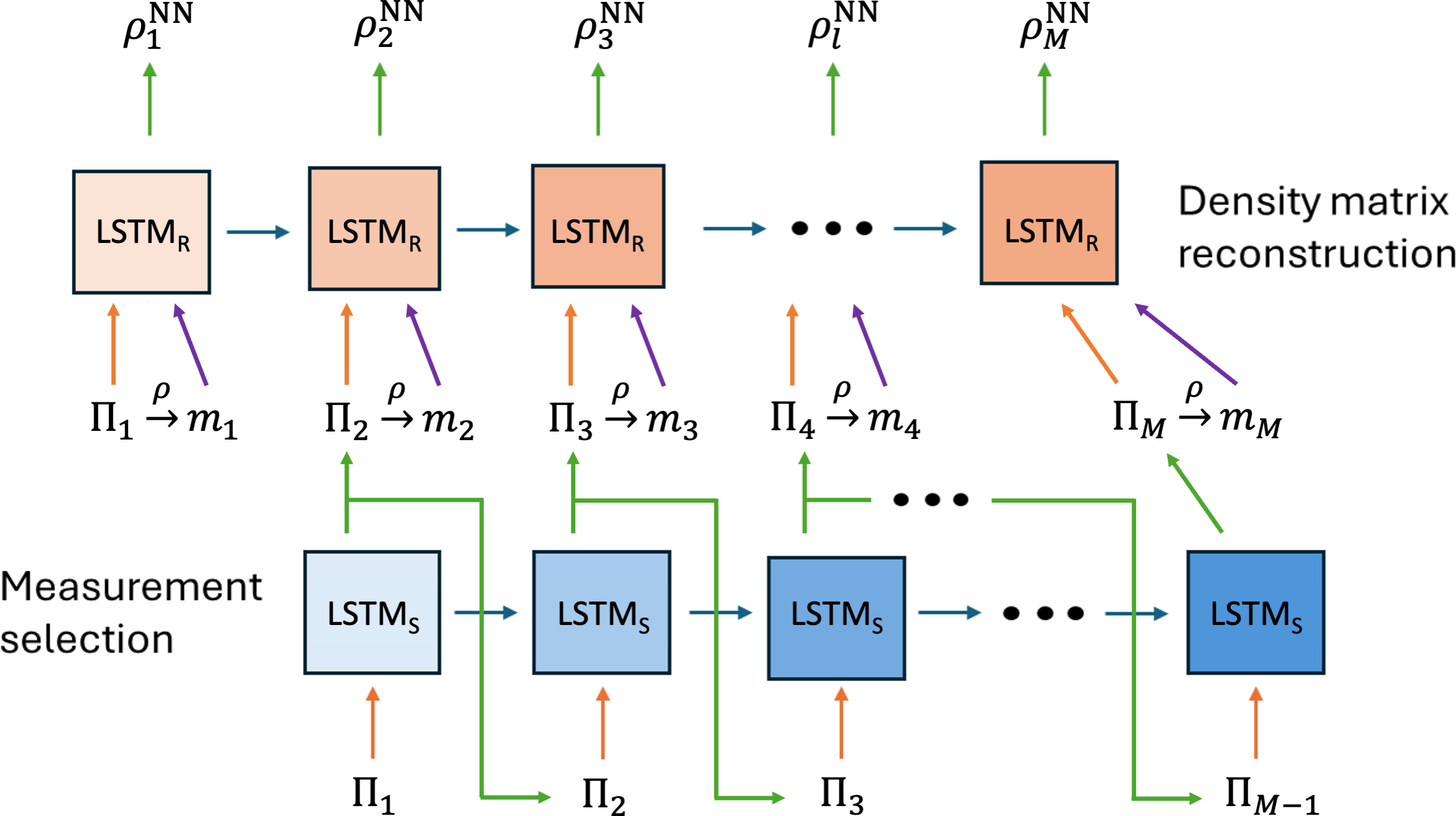}
    \caption{Scheme for the neural-based measurements selection and reconstruction network. Having a previous selection $\Pi_{l-1}$, the goal for the recurrent autoregressive LSTM$_\mathrm{S}$ unit is to propose the next measurement operator $\Pi_l$, which is used to calculate the value $m_l$. Simultaneously, a pair $(\Pi_l,m_l)$ is used as an input to LSTM$_\mathrm{R}$ cell, which outputs increasingly better reconstructions $\rho_l$, where $l$ denotes the number of already performed measurements.}
    \label{fig:nn_selection_with_reconstruction}
\end{figure}

\begin{figure*}[t]
\centering
\includegraphics[width=1.\linewidth]{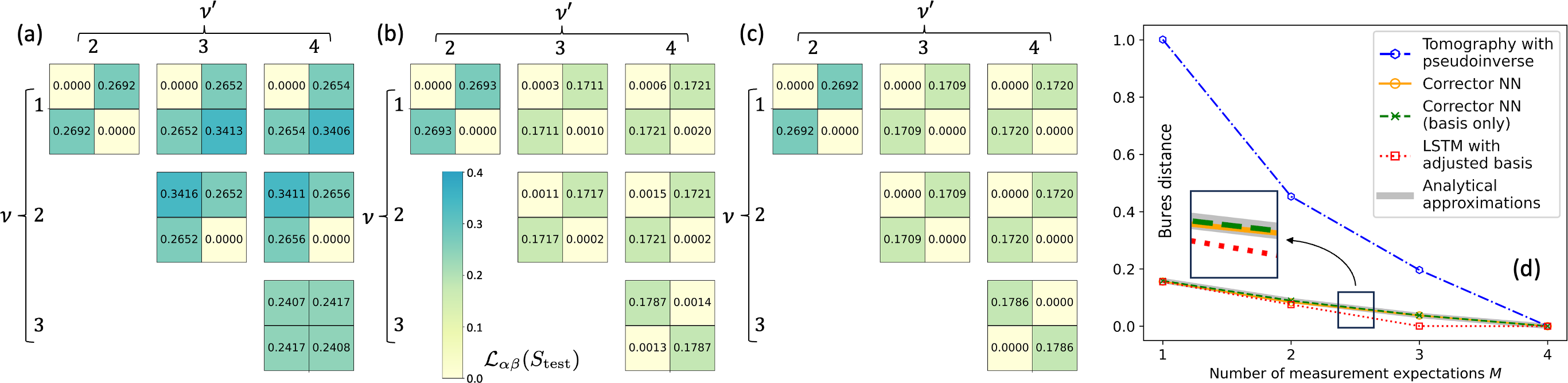}
\caption{Undercomplete 1-qubit density matrix tomography with two measurements $(\nu,\nu')$. (a-c) Element-wise reconstruction error $\mathcal{L}_{\alpha\beta}(S_\mathrm{test})$ using: (a) pseudoinverse reconstruction, (b) NN corrector, (c) best possible analytical formula. (d) Bures distance to target matrix averaged over $S_\mathrm{test}$ for the different techniques with increasing number of measurements. \bluemark{All panels show errors of the raw reconstruction outputs. No trace normalization or eigenvalue clipping is applied before comparison.}}
    \label{fig:1_qbit}
\end{figure*}

\section{Neural models}
In the first attempt, we train a NN just to correct the pseudoinverse $B^+_{\mu\nu}$ matrix by adding some correction term $b^\mathrm{NN}_{\mu\nu}$ in order to get the smallest possible reconstruction error averaged over the training set: 
\begin{equation}
\rho^\mathrm{NN}_\mathrm{corr}(\mathcal{M})=\Gamma_\mu\!\left(\left(B^+_{\mu\nu}+b^\mathrm{NN}_{\mu\nu}(\mathcal{M})\right)m_\nu+c^\mathrm{NN}_\mu(\mathcal{M})\right),
\label{eq:corrector}
\end{equation}
where \bluemark{an incomplete measurement set $\mathcal{M}=\{(\Pi_\nu)_{\nu=1}^M,(m_\nu)_{\nu=1}^M\}$ is a subset sampled without replacement from the catalog ${\cal C}_N$ in Eq.~\eqref{eq:projector_catalogue}; 
while $c^\mathrm{NN}_\mu$ is added to enable the model to predict normalized $\rho^\mathrm{NN}$, irrespectively of the $m_\nu$ values.
Thus, for each value of $M$, the measurement subset is random, but contains no repeated projectors.}
The network input, therefore, contains both the selected projectors and, depending on the model variant, \bluemark{the corresponding measurement expectation values. For each $M$, we train and test the network for 10 independently sampled subsets and report the average reconstruction performance. This procedure is used to characterize the average behavior of the reconstruction map for undercomplete data, rather than the performance for one specially chosen measurement sequence.}
We call this model the \textit{corrector} network, and its details are discussed in the Appendix~A.

\bluemark{
We also introduced a modified corrector version $\bar{\rho}^\mathrm{NN}_\mathrm{corr}(\Pi)$ that predicts coefficients using only the information about measurement operators (that do not depend on measurement expectation values) -- defined in the same way as in Eq.~\ref{eq:corrector} but now $\mathrm{NN}_\mathrm{corr}(\Pi)=(b^\mathrm{NN}_{\mu\nu}(\Pi),c^\mathrm{NN}_\mu(\Pi))$, where we abbreviated $\Pi\equiv(\Pi_\nu)_{\nu=1}^M$.
One can also imagine improved version of the corrector NN, i.e. one containing a \textit{quadratic} term:
\begin{equation}
\bar{\rho}^\mathrm{NN}_\mathrm{M^2\mhyphen{}corr}(\Pi)=\bar{\rho}^\mathrm{NN}_\mathrm{corr}(\Pi)+\Gamma_\mu S^\mathrm{NN}_{\mu\nu\nu'}(\Pi)\, m_\nu{}m_{\nu'},
\label{eq:corrector2}
\end{equation}
with the network $S^\mathrm{NN}_{\mu\nu\nu'}(\Pi)$ architecture designed in a way that the output tensor $S_{\mu\nu\nu'}$ is symmetric -- invariant under any measurements ($\nu$ and $\nu'$) permutation.} 

NNs may become useful not only for reconstructing the density matrix from the reduced number of measurement expectations, but also for selecting the most important measurements~\cite{Balaz2025}. The general idea is to use two separate \bluemark{long short-term memory (LSTM)} networks presented in Fig.~\ref{fig:nn_selection_with_reconstruction}, where the first one, LSTM$_\mathrm{S}$, chooses the order of the measurements, and the second, LSTM$_\mathrm{R}$, exploits the information contained in the measurements to reconstruct the density matrix effectively.
\bluemark{At step $l$, the selected by LSTM$_S$ pair $(\Pi_l,m_l)$ is passed to LSTM$_R$, which returns an intermediate reconstruction $\rho_l^{\rm NN}$. The reconstruction error of $\rho_l^{\rm NN}$ is then used in the training loss. In this way, LSTM$_S$ is trained indirectly through the quality of the reconstruction produced by LSTM$_R$. 
%There is no additional explicit feedback connection in which $\rho_l^{\rm NN}$ itself is used as an input to LSTM$_S$ for choosing $\Pi_{l+1}$. Instead, the selector bases its next choice on its recurrent hidden state, which stores information about the previously chosen measurements.
More details about the \textit{selector} model can be found in the Appendix~A}.

\bluemark{We compare three sequential scenarios. In the first one, called LSTM with \emph{random basis sequence}, only the reconstruction network LSTM$_R$ is used: the projectors are chosen randomly without replacement from ${\cal C}_N$, and LSTM$_R$ reconstructs the state from the resulting sequence. This provides a baseline showing the effect of recurrent reconstruction without learned measurement selection.
In the second scenario, called LSTM with \emph{predefined} (James \emph{et al.}) \emph{basis}, LSTM$_S$ chooses projectors from the discrete catalog ${\cal C}_N$, again without repetitions.
In the third scenario, called LSTM with \emph{adjusted basis}, LSTM$_S$ is not restricted to the discrete catalog and instead outputs a product projector directly. We refer to these projectors as adjusted measurement operators: they are still valid product projectors, but their local measurement directions need not coincide with the catalog in Eq.~\eqref{eq:projector_catalogue}.}

\bluemark{The overall numerical procedure is as follows. First, training and test sets of random density matrices, $S_{\rm train}$ and $S_{\rm test}$, are generated as described in Appendix~B. For the corrector networks and a fixed value of $M$, we sample incomplete measurement sets ${\cal M}\subset{\cal C}_N$ without replacement, compute the exact expectation values $m_\nu=\mathrm{Tr}(\Pi_\nu\rho)$, and train the network parameters by minimizing the mean $L_2$ reconstruction error over $S_{\rm train}$. For the sequential LSTM models, the same reconstruction loss is accumulated over all steps $l=1,\ldots,M$: LSTM$_S$ selects the measurement projectors, LSTM$_R$ reconstructs the density matrix after each new measurement, and the two modules are optimized jointly. Finally, to describe the fidelity of a state reconstruction, the trained models are evaluated on $S_{\rm test}$ using the so-called Bures distance~\cite{bengtsson2017,Gilchrist2005}:}
\begin{equation}
    %\mathcal{B}(\rho^\mathrm{NN},\rho)=\sqrt{2-2\sqrt{F(\rho^\mathrm{NN},\rho)}},
    \mathcal{B}(\rho^\mathrm{NN},\rho)=2-2\sqrt{F(\rho^\mathrm{NN},\rho)},
    \label{eq:Bures}
\end{equation}
with the \textit{fidelity} $0\leq F \leq 1$ defined as
\begin{equation*}
F(\rho^\mathrm{NN},\rho)=\Tr\!\left(\sqrt{\sqrt{\rho^\mathrm{NN}}\rho\sqrt{\rho^\mathrm{NN}}}\right)^2\!.
\end{equation*}
\bluemark{Here $\rho^{\rm NN}$ denotes the direct output of the considered reconstruction method. No additional physical projection is applied before evaluating the error. When a method produces a non-normalized or non-positive operator, this unphysicality contributes to the increased reconstruction error. The quantity should therefore be understood as a raw-output Bures-type score rather than as a post-processed physical-state distance.}
\bluemark{To accumulate statistics for different values of $M$, the whole procedure is then repeated for $M\in\{1, ..., 4^N\}$}

\section{1-qubit results}
As the first step of the evaluation, we verify the proposed approach for the trivial 1-qubit system.
In Fig.~\ref{fig:1_qbit} we present the reconstruction error for a 1-qubit density matrix, where we calculate the average error element-wise: $\mathcal{L}_{\alpha\beta}(S_\mathrm{test})=\frac{1}{N_\mathrm{test}}\sum_i|\rho_{i,\alpha\beta}-\rho_{i,\alpha\beta}^\mathrm{NN}|$, $\rho_i\in{}S_\mathrm{test}$.
We plot the results for different pairs of measurement operators $(\nu,\nu')$.
One can notice that the NN corrector, as presented in Fig.~\ref{fig:1_qbit}(b), has almost the same structure of errors as the analytical formulas -- Fig.~\ref{fig:1_qbit}(c), which give the best possible reconstruction for a given pair of measurements. These formulas can be easily found in a 1-qubit system and are discussed in Appendix~C. What is characteristic, the optimal selection of $b^\mathrm{NN}_{\mu\nu}$ and $c^\mathrm{NN}_\mu$ does not depend on the measurement results $m_\nu$ but only on the measurement operators $\Pi_\nu$. Therefore, $\rho^\mathrm{NN}_\mathrm{corr}(\mathcal{M})$ (as well as $\bar{\rho}^\mathrm{NN}_\mathrm{M^2\mhyphen{}corr}(\mathcal{M})$) simply collapses to $\bar{\rho}^\mathrm{NN}_\mathrm{corr}(\Pi)$. Measurement expectations $m_\nu$ become relevant for the reconstruction of the 2-qubit and larger systems.
It is worth noting that the optimal reconstruction found by NN outperforms an approximate one using the pseudoinverse -- Fig.~\ref{fig:1_qbit}(a). The latter often has problems with preserving the trace-norm, as discussed in Appendix~C.

Similar error precedence can be observed when examining the ($S_\mathrm{test}$-averaged) reconstruction (error) measure $\langle \mathcal{B}\rangle$, for different numbers of measurements of a single qubit. Fig.~\ref{fig:1_qbit}(d) shows that the Bures distance for various QST techniques decreases with the number of measurements and approaches zero when $M = 4$. The reconstruction obtained with the NN corrector (orange curve) is much better than pseudoinverse (blue) and practically as good as (the best) analytical approximation (gray), which is further discussed in Appendix~C.
The NN corrector that takes only information about the basis, i.e. $\Pi_\nu$, (green) performs similarly.
The LSTM model (red) additionally optimizes the measurement order, avoiding redundancy, and obtains a complete reconstruction just with $M=3$ measurements, that for 1-qubit setup with optimized choice of measurements (i.e. one that utilizes trace-normalization -- see Appendix~C) are sufficient.
\bluemark{Note that if the measurement set is not optimal (i.e., is redundant), then even three measurements do not provide complete information about the state. Consequently, even in the best analytical reconstruction, the gray curve in Fig.~\ref{fig:1_qbit}(d) exhibits a nonzero error for $M=3$.}  

\section{2-qubit results}
\begin{figure}[tb]
    \includegraphics[width=.99\linewidth]{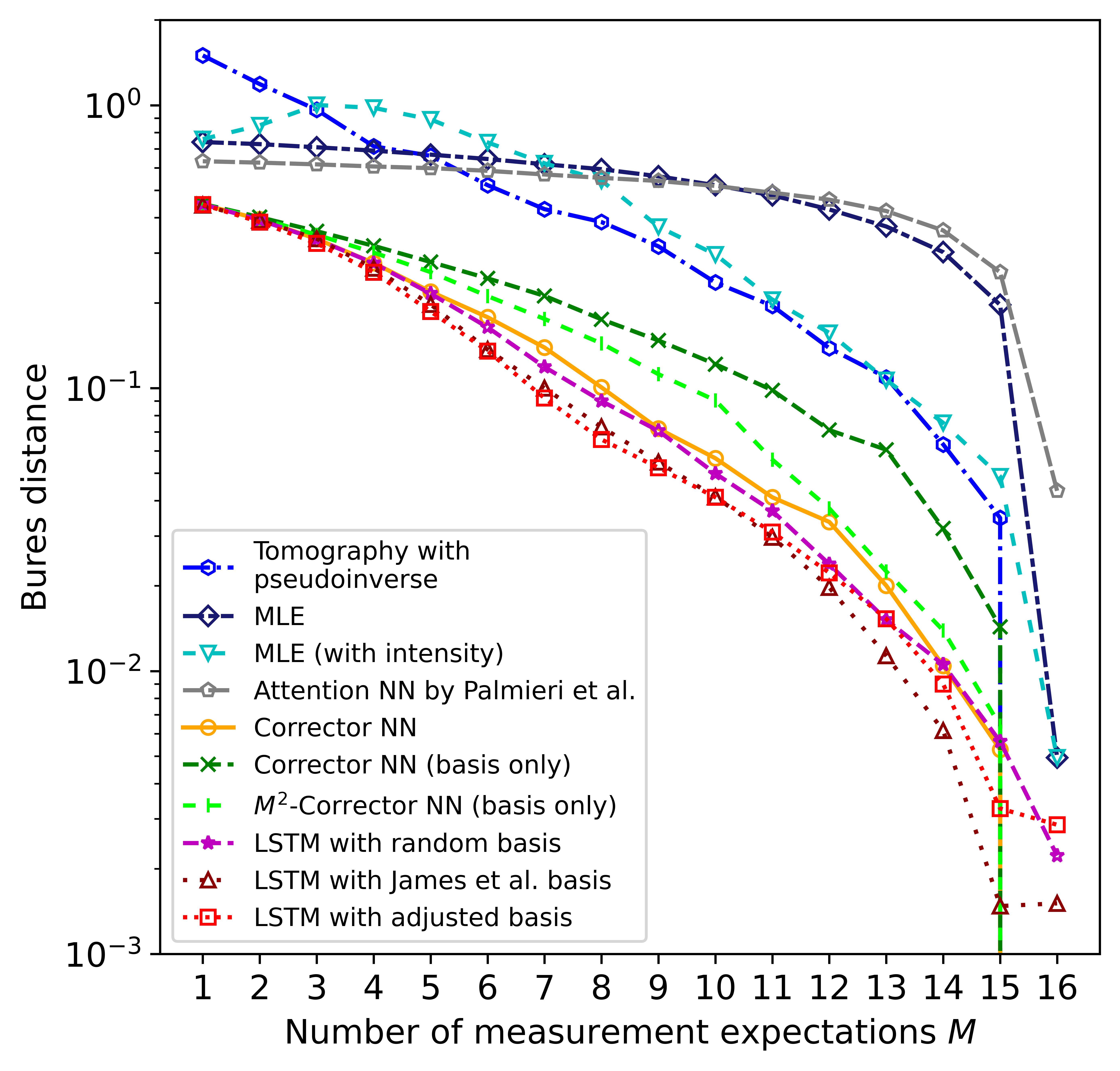}
    \caption{Undercomplete tomography of 2-qubit random mixed states: $S_\mathrm{test}$-averaged Bures distance $\langle\mathcal{B}\rangle$ to target matrix for the different reconstruction techniques with increasing number of measurement outcomes $M$.}\label{fig:2_qbit}
\end{figure}

Let us now move to a 2-qubit system, whose geometry is far more nontrivial~\cite{Omar2016,Jakubczyk2001,Kus2001} and even describing semi-positivity is nontrivial itself.
The average Bures distance $\langle\mathcal{B}\rangle$ vs. the number of measurements $M$ for different reconstruction approaches for the set $S_\mathrm{test}$ of random mixed states is presented in Fig.~\ref{fig:2_qbit}.
Although all NN models achieve far better scores in density matrix reconstruction than the pseudoinverse-based QST (blue), they also outperform the MLE method-based reconstructions (computed using {\tt Quantum-Tomography} library~\cite{quantum-tomography}) both when calculated with (light blue) and without (dark blue) the \textit{intensity} parameter. It informs the MLE which basis measurements were present (with intensity 1) or absent (with zero intensity) during the fitting process.
Our protocols also outperform the attention-based NN QST (LI-NN model) introduced by Palmieri \textit{et al.}~[\onlinecite{Palmieri2024}] (grey curve).
\bluemark{In the LI-NN model, the attention mechanism assigns larger weights to the most informative measurement expectation values, enabling the NN to more efficiently extract the information relevant for QST. To adapt the method to our setting, we add substantially stronger Gaussian noise (with 100 times larger standard deviation, $\sigma=0.158$) to those measurement expectation values that are not used in a given test (i.e., those that are \textit{out of the box}). This differs from the original methodology~\cite{Palmieri2024} in which variable noise levels were added uniformly to \textit{all} measurements.}

When it comes to the NN corrector model, it is clear that including information about the measurement values in $\rho^\mathrm{NN}_\mathrm{corr}(\mathcal{M})$ (orange), helps the NN to predict slightly more accurate corrections to the pseudoinverse method than $\bar{\rho}^\mathrm{NN}_\mathrm{corr}(\Pi)$ (dark green). 
%One can also observe that including the quadratic term in 
One can also observe that the inclusion of the quadratic term in
$\bar{\rho}^\mathrm{NN}_\mathrm{M^2\mhyphen{}corr}(\Pi)$ (light green), Eq.~\ref{eq:corrector2}, gives better reconstruction than the linear model (dark green), but still worse than the measurement value-dependent one (orange).
%which suggests that the optimal reconstruction theory might include even higher-order measurement interactions.

% Moreover, there is a visible improvement in performance for the LSTM recurrent networks, which are not only tasked with reconstructing (LSTM$_\mathrm{R}$ network), but also to propose a sequence of measurement operators (LSTM$_\mathrm{S}$): the brown curve, in Fig.~\ref{fig:2_qbit}, presents Bures distance $\langle\mathcal{B}\rangle$ for the model that selects measurement operators from a predefined basis, Eq.~(\ref{eq:single_qubit_catalogue}), while the red curve evaluates the model that proposes custom operators. The performance gain from adjusting the sequence $(\Pi_\nu)$ suggests that some measurements are of greater importance in the $S_\mathrm{train}$ set, which we discuss further.
% An important observation is that the predefined basis \textit{selector} (brown) collapses to a single measurement sequence already during the training (albeit different trainings lead to different sequences); while the custom basis-\textit{selector} (red) during tests follows various paths in the $\Pi_l$ operator space, which proves that there are many equivalent possibilities for selecting optimal sequences of measurement bases.
% For comparison, the results for the LSTM model that reconstructs a density matrix using a random operator basis, i.e., composed only of LSTM$_\mathrm{R}$ (the upper orange part in the scheme in Fig.~\ref{fig:nn_selection_with_reconstruction}) are shown as the violet curve in Fig.~\ref{fig:2_qbit}. 

\bluemark{The sequential LSTM results show the benefit of learning the measurement order. The violet curve in Fig.~\ref{fig:2_qbit} corresponds to the baseline LSTM with a random basis sequence: only LSTM$_R$ is used, while the measurement projectors are drawn randomly without replacement from ${\cal C}_N$. The brown curve corresponds to LSTM with predefined basis, where LSTM$_S$ learns to choose projectors from the catalog ${\cal C}_N$. In this case, after training, the selector typically assigns almost all probability to one ordered list of catalog projectors. In other words, we can say that the selector ``collapses'' to a single measurement sequence. However, different training runs may still converge to different, nearly equivalent sequences. The red curve corresponds to the LSTM with adjusted measurement basis, where LSTM$_S$ is allowed to output product projectors not restricted to the discrete catalog. In this case, the learned measurement directions can depend more continuously on the previous measurement history, and the trajectories in projector space need not coincide with a fixed list of catalog elements.}

Also in Fig.~\ref{fig:2_qbit}, we can observe that for the complete measurement set ($M=16$), the corrector schemes exhibit no error at all (the correction is trivial, no correction is needed), while the LSTM models that reconstruct a state from scratch show a small but finite error. From this perspective, the LSTM with predefined basis (brown curve) performs better.
%Kolejnosc przy LSTM-Kwiat (ustalona na etapie treningu): [0,11,14,1,9,13,3,6,10,7,2,5,15,12,8,4]. Inny trening: [0,9,1,14,15,12,13,2,5,3,7,6,8,10,11,4]. Jeszcze inna [0,13,9,10,3,1,2,5,8,6,11,14,12,4,7,15].
%On the other hand, the choice of the strategy for the measurement selection seems to be less important, however, the model without any constraints seems to perform slightly better than the one that is forced to choose from the discrete basis of measurement set.

\begin{figure}[b]
    \centering
    \vspace{-3mm}\includegraphics[width=.99\linewidth]{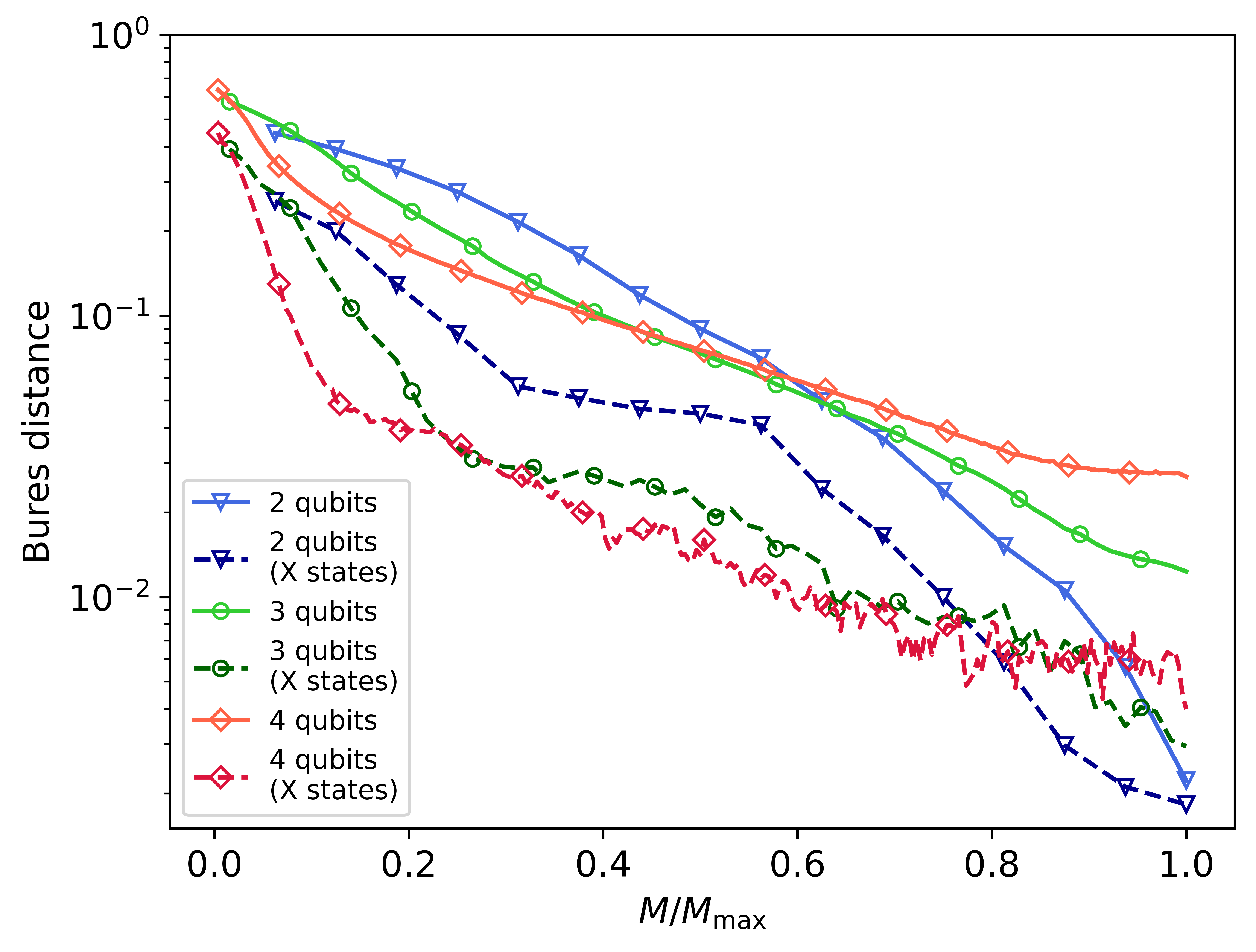}
    \caption{Undercomplete tomography for restricted states: average Bures reconstruction distance $\langle\mathcal{B}\rangle$ as a function of the number of measurement outcomes $M$ ($M_\mathrm{max}=4^N$), for  model trained on general mixed states (solid lines) and on the restricted X-states family (dashed lines) across different system sizes $N=2,3,4$. Results for LSTM with random basis.}
    \label{fig:scaling_x_states}
\end{figure}

\begin{figure*}[tb]
    \centering
    \includegraphics[width=.9\linewidth]{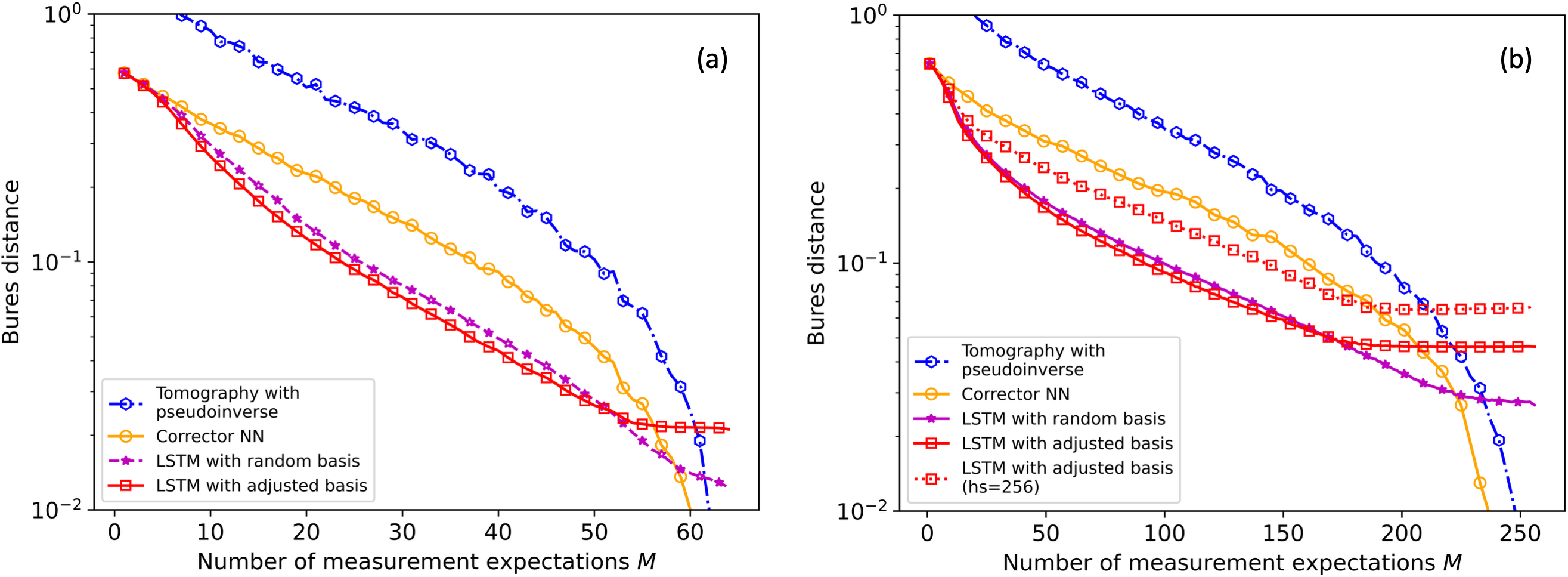}
    \caption{Undercomplete (a) 3-qubit and (b) 4-qubit tomography of random mixed states: average Bures distance $\langle\mathcal{B}\rangle$ to a target matrix for the different reconstruction techniques with increasing number of measurement expectation values $M$: (a) $M=1,\dots,4^3$, and (b) $M=1,\dots,4^4$.
    Additional dotted red curve in (b) presents results for the smaller LSTM with adjusted basis model.}
    \label{fig:3qbits_results}
\end{figure*}

In general, for an arbitrary state, there is no predefined optimal sequence of measurements, which are linearly independent operators. However, if we restrict the state space to a certain subclass, e.g. to so-called X states~\cite{Quesada2012}, certain sequences of measurements are preferable to others. Typically, the first three gave full information about the diagonal elements; while the next four gave full information about the coherences. It is further discussed in Appendix~D. In Fig.~\ref{fig:scaling_x_states}, we compare the performance of the LSTM model when trained on a restricted space of X states versus training on general mixed states (as in Fig.~\ref{fig:2_qbit}).
We observe that the model can adapt when trained on restricted states.
Increasing the system size also affects the model's performance. In particular, for X states, where the number of effective parameters scales as $2^{N+1}$ compared to $4^N-1$ for general mixed states, it becomes evident that significant information about the quantum state accumulates in the initial measurements and can be efficiently learned by the NN.

\section{Scaling up to larger systems}

All the presented methods can be scaled to larger systems -- up to 3 (4) qubits with $4^3=64$ ($4^4=256$) measurements, as presented in Fig.~\ref{fig:3qbits_results}(a) and (b), respectively. These results illustrate tomographic reconstruction for random mixed states of 3 and 4 qubits. Comparing them with the 2-qubit results shown in Fig.~\ref{fig:2_qbit}, we observe that, as before, our proposed protocols significantly outperform standard methods based on the pseudoinverse; however, it is worth adding that the pseudoinverse reconstruction decreases with the number of measurement expectation values $M$ in the same exponential way as the ML-based methods.
We also see that the results scale qualitatively: all curves exhibit a similar character in the plots for 3 and 4 qubits.
Two characteristic features can also be observed: LSTM-based methods perform slightly better than the corrector NN; on the other hand, for larger systems, a small but finite error appears in the LSTM methods toward the end of reconstruction, when the measurement basis is nearly complete. This is related to the fact that recurrent networks (even LSTMs) may ``forget'' the earliest measurement expectations. 
%Addressing this issue will be the subject of future work.

\begin{figure}[b]
    \centering
    \includegraphics[width=.95\linewidth]{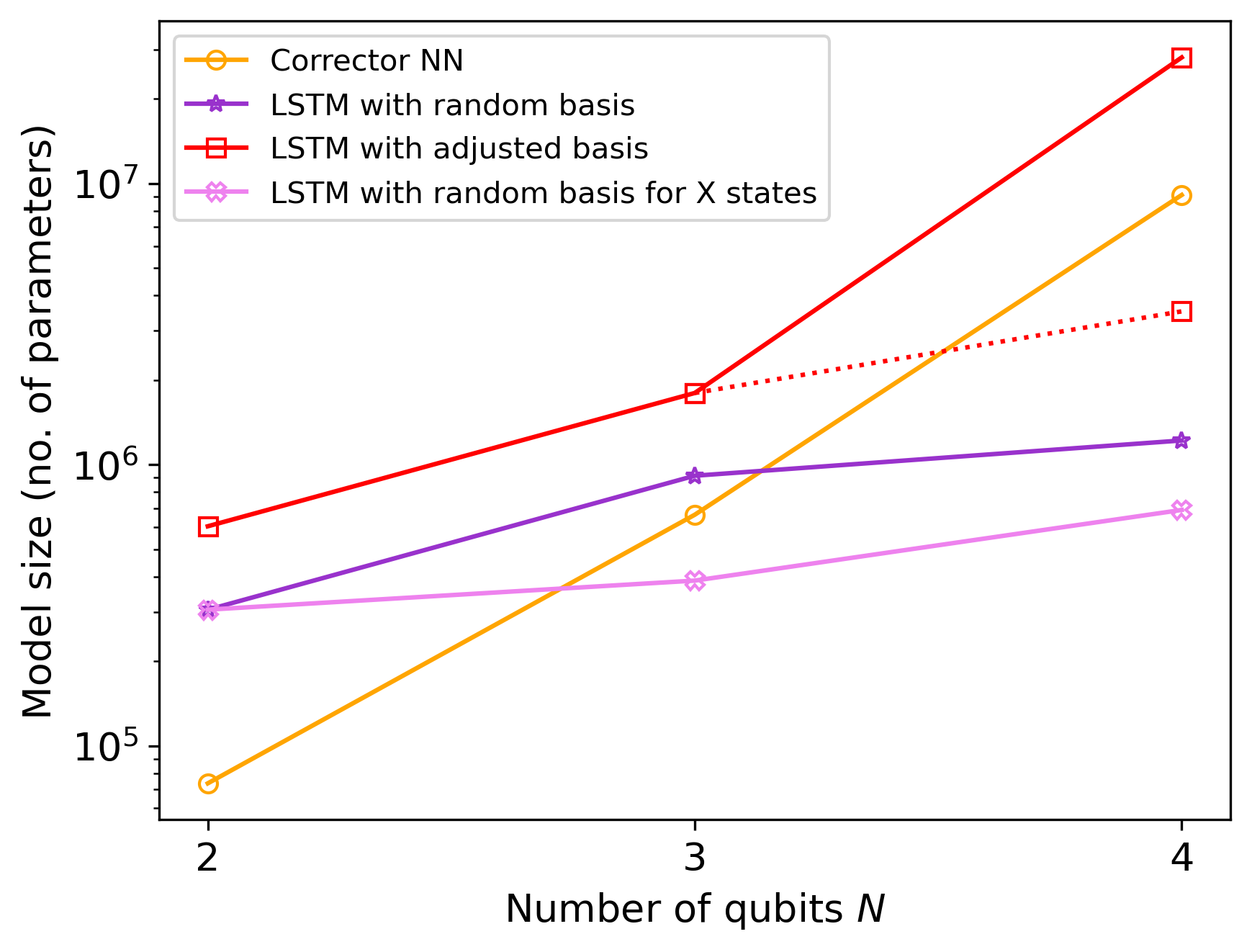}
    \caption{Scaling of the analyzed neural network models with the system size $N$. Additional dotted red curve presents scaling for the smaller LSTM with adjusted basis model.}
    \label{fig:param}
\end{figure}

Regarding the scaling of model size, which determines their capacity when the system size $N$ changes -- at each size, the capacity was chosen to ensure comparable performance between models at different system sizes. 
The criterion was to achieve a Bures distance about 5 times smaller at the halfway measurement range ($M=4^N\!/2$) for the LSTM-based models compared to the pseudoinverse.  
For LSTM models, we had to increase the number of layers from 1 to 2 for 3 and 4-qubit systems. Additionally, for LSTM with adjusted basis, to keep the satisfactory level of performance for 4 qubits, we further modified the hidden size to 1024 units. 
In the case of the corrector NN, the increase in the number of parameters follows directly from the input and output scaling, while the architecture itself was kept unchanged.

\begin{table}[t]
\centering
\begin{tabular}{lccc}
\hline
\textbf{Model} & \textbf{2 qubits} & \textbf{3 qubits} & \textbf{4 qubits} \\
\hline
Corrector NN & \(7.36 \times 10^{4}\) & \(6.64 \times 10^{5}\) & \(9.11 \times 10^{6}\) \\
LSTM with random basis            & \(3.05 \times 10^{5}\) & \(9.14 \times 10^{5}\) & \(1.22 \times 10^{6}\) \\
LSTM with adjusted basis  & \(6.04 \times 10^{5}\) & \(1.80 \times 10^{6}\) & 
\begin{tabular}{@{}c@{}}\(2.81 \times 10^{7}\) \\ \(( 3.51\times 10^{6})\)\end{tabular}\\
%\(2.81 \times 10^{7} 3.51 \times 10^{6}\) \\
\hline
\end{tabular}
\caption{Number of neural network parameters for different models and system sizes. \bluemark{The number of parameters for LSTM with adjusted basis for 4 qubits is presented both for the model with 1024 hidden units and 256 units (in brackets).}}
\label{tab:param}
\end{table}
The total number of learnable parameters in the analyzed models are listed in Table~\ref{tab:param} and illustrated in Figure~\ref{fig:param}. From Fig.~\ref{fig:param}, we see that as the system size $N$ increases, the capacity of the NN models must also be increased. The increase of the NN corrector (orange curve) is exponential and is directly related to the change in the dimension $2^N$ of the state space. For both LSTM-based models (red and violet curves), their size also grows exponentially, which is natural since we tested them on general mixed states. For example, in the case of LSTM with an adjusted basis for $N=4$ qubits -- shown as the solid red curve in Fig.~\ref{fig:3qbits_results}(b) -- we intentionally increased the number of hidden units in the LSTM cell to 1024 to achieve a performance comparable to that of $N=2,3$. If instead we keep the number of hidden units the same as for $N=2,3$ qubits (i.e., hidden size $\mathrm{(hs)}=256$) -- giving the scaling shown in Fig.~\ref{fig:param} by the dotted red curve -- the performance decreases, as indicated by the dotted red curve in Fig.~\ref{fig:3qbits_results}(b).
Extrapolating the observed parameter count shown in Fig.~\ref{fig:param} suggests that analogous architectures could be constructed for larger systems, with the parameter count remaining below $\!10^9$ up to approximately 6 qubits for the architectural choices considered here.

\section{Positive semidefiniteness}
\label{sec:semidefiniteness}
\begin{figure}[b]
    \includegraphics[width=.99\linewidth]{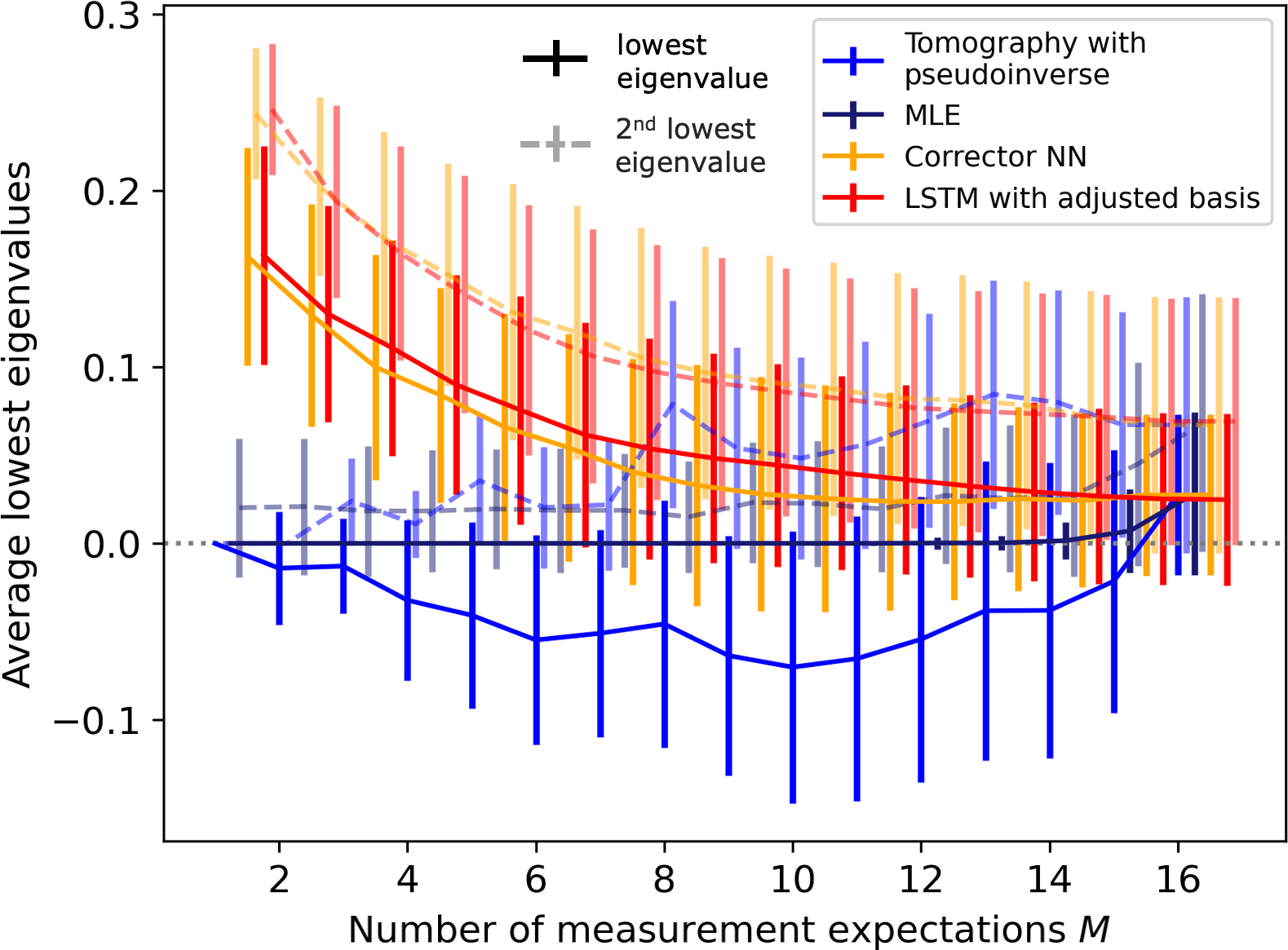}
    \caption{Positive semidefiniteness of density matrices reconstructed using different techniques. Shown are the average lowest (intense color) and second-lowest (faded color) eigenvalues, with error bars indicating the standard deviation.}\label{fig:pos}
\end{figure}
As additional tests for the models' performance, we also verified that both the corrector NN and the LSTM models reconstruct a positive semidefinite matrix: regardless of the measurement collection size $M$, the reconstructed matrix has the lowest eigenvalue that is on average always nonnegative, as presented in Fig.~\ref{fig:pos}. Here the two lowest eigenvalues: 1\textsuperscript{st} and 2\textsuperscript{nd} (marked as an intense and faded color, respectively) are represented as the average eigenvalue and its standard deviation (marked as an error bar).
This shows that NNs tend to preserve the geometry of mixed states, which is not true for the pseudoinverse method.
%A characteristic feature of MLE is that it pins the lowest eigenvalue to zero by default, even at the cost of poor reconstruction.
\bluemark{MLE often returns rank-deficient density matrices whose smallest eigenvalues are exactly zero because the optimum lies on the boundary of the physical state space}~\cite{Blume2010}.
\bluemark{We stress that the positivity analysis shown here is diagnostic only. Positive semidefiniteness is not imposed on the reconstructed matrices during evaluation. Instead, the eigenvalues are computed directly from the raw outputs of the corresponding reconstruction methods.}

\section{Discussion}
The results presented here demonstrate that learning-based approaches can significantly improve QST for informationally incomplete measurement data. Across all tested system sizes, NN estimators consistently outperform standard reconstruction techniques, including pseudoinverse-based inversion and MLE, even when trained and evaluated on general mixed states. \bluemark{Results for 3- and 4-qubit random mixed states show that the qualitative behavior observed for two qubits persists for larger systems: the NN-based methods remain substantially more accurate than the pseudoinverse reconstruction in the undercomplete regime, and the LSTM-based schemes typically outperform the corrector network at intermediate values of $M$. At the same time, the larger-system results reveal a limitation of the recurrent approach. Close to the complete-measurement regime, the LSTM reconstructions may retain a small but finite error, even when almost all measurement expectation values are available. We attribute this effect to the finite memory of recurrent architectures: despite the gating mechanism of LSTM cells, information from the earliest measurements can be partially forgotten after a long sequence of later inputs. This suggests that future implementations may benefit from architectures with more direct access to the full measurement history, such as attention-based or transformer-like models.}

Two complementary strategies were explored: (1) learning optimal linear or quadratic reconstruction maps conditioned on the chosen measurement operators; (2) pair of recurrent LSTM networks performing sequential reconstruction while adaptively selecting subsequent measurements. In both cases, the networks can effectively learn the geometry of quantum states, e.g., positive semidefiniteness.

%While the learning protocols provide improvements already in the general case, their advantages become particularly pronounced when prior information about the structure of the quantum state space is available. 
Training on restricted families of states leads to faster convergence and reveals preferred measurement sequences, indicating that the networks can adaptively tailor the reconstruction protocol to the relevant region of the state space. In this sense, the results provide an operational manifestation of a \emph{no-free-lunch} principle for QST: efficiency gains are achievable when some assumptions of the underlying state space can be learned.

%The present approach does not circumvent the exponential scaling inherent to full QST. Any method aiming to reconstruct arbitrary mixed states must ultimately contend with the growth of the state space dimension as $4^N-1$. Instead, the learning-based framework explored here reallocates complexity from measurement completeness to prior specification: when the experimentally accessible states occupy a restricted subspace of the full Hilbert space, NNs can leverage this structure to reduce the number of required measurements and to prioritize informationally relevant observables.

Learning-assisted QST complements existing approaches such as compressed sensing~\cite{Gross2010}, Bayesian inference, and shadow tomography. Whereas these methods rely on analytically specified priors or targeted observable prediction, NNs offer a general and adaptable framework capable of learning reconstruction protocols directly from measurement statistics. The results presented here suggest that such methods can enhance state characterization in both general and structured scenarios, particularly in experimentally relevant regimes where measurement resources are limited.

\bluemark{A limitation of the present numerical analysis is that it is performed in the infinite-shot regime. For a finite number $N_c$ of copies, each input probability $m_\nu$ should be replaced by an empirical frequency $\hat m_\nu=n_\nu/N_c$, with binomial variance $m_\nu(1-m_\nu)/N_c$. %for an ideal binary projective measurement. 
Such sampling noise would propagate through both the pseudoinverse estimator and the neural reconstructions, and its quantitative impact is expected to depend on $N_c$, the selected projectors, and the conditioning of the reconstruction map. The present work deliberately separates this effect from the error due to the explicit focus on the informational incompleteness scenario. Incorporating realistic finite-shot fluctuations, detector noise, and possible noise-aware training or data augmentation is a natural next step toward experimental deployment of the proposed schemes.}

\section*{Acknowledgments}
MK and JP acknowledge support from National Science Centre, Poland, under grant no. 2021/43/D/ST3/01989. 
KR is funded within the QuantERA II Programme that has received funding from the EU H2020 research and innovation programme under Grant Agreement No.~101017733, and with funding organization MEYS.
This work was supported by the research programme of the Strategy AV21 AI: Artificial Intelligence for Science and Society.
We gratefully acknowledge Polish high-performance computing infrastructure PLGrid (HPC Centers: ACK Cyfronet AGH) for providing computer facilities and support within computational grant no. PLG/2024/017284.

\section*{Data availability}
The datasets generated and analyzed during the current study are available in the Zenodo repository, \url{https://doi.org/10.5281/zenodo.17276582}.

%\section*{Competing interests}
%All authors declare no financial or non-financial competing interests.

%\section*{Author contributions}
%KR and JP conceived the project idea. JP and MK designed the neural network architectures. MK and PB conducted the numerical experiments, including model training and evaluation. All authors analyzed and discussed the results. All authors contributed to writing and approved the final manuscript.

\appendix
\section{Neural models}
\subsection{Corrector network details} 
The \textit{corrector} NN, introduced in the Eq.~\ref{eq:corrector} of the main text, has two outputs: $\mathrm{NN}_\mathrm{corr}(\mathcal{M})=(b^\mathrm{NN}_{\mu\nu}(\mathcal{M}),c^\mathrm{NN}_\mu(\mathcal{M}))$. The second output, $c^\mathrm{NN}_\mu$ vector, is conceived to enable the model to predict normalized $\rho^\mathrm{NN}$, irrespectively of the expectation values $(m_\nu)_{\nu=1}^M$, $M<4^N$ (which no longer form a complete tomographic set). Moreover, if we take Eq.~\ref{eq:corrector} and calculate element $\langle\psi_{\nu'}\!|\,.\,|\psi_{\nu'}\!\rangle$ of both sides, we arrive at $B_{\nu'\mu}v_\mu=0$, $v_\mu=b^\mathrm{NN}_{\mu\nu}m_\nu+c^\mathrm{NN}_\mu$ for all $\nu'=1...M$ (see Appendix~E for derivation), which means that the two NN outputs are not independent, and vector $v_\mu$ must belong to the orthogonal complement of the subspace spanned by rows of $B_{\nu'\mu}$. To fulfill this condition, we added an extra component to the corrector NN loss during its training. 
Although the orthogonality condition speeds up training, it is not necessary: if we remove it, the corrector NN learns it anyway.
It turns out that in the 1-qubit case, i.e., $N=1$, taking an appropriate (depending on the measurement basis $(\Pi_\nu=\mu_{i_1})_{i_1=1}^{M<4}$) orthogonal vector $v_\mu$ (from the orthogonal complement) and choosing its norm so that $\Tr(\rho^\mathrm{NN})=1$ is equivalent to taking the best possible analytical formula for the reconstruction, as discussed in the Appendix~C. However, for two and more qubits the choice is not so obvious anymore.
Also, for a single qubit, the optimal $b^\mathrm{NN}_{\mu\nu}$ and $c^\mathrm{NN}_\mu$ do not depend on the results of measurements $m_\nu$ (we can simply remove this information from training, getting the same results), which is not the case for $N>1$.
%due to the fact that qubits can be correlated. 
%\comment[magenta]{Zauwazmy ze jezeli b i c nie zaleza od m to znaczy ze sa stale dla danego wyboru bazy pomiarów}

%To speedup the learning process, separate trainings were performed for different measurement operators $(\Pi_\nu)_{\nu=1}^M$. These models can be combined into one NN \textit{corrector} to which the measurement basis information will be input, as stated in Eq.~\ref{eq:corrector}.

Note that the input layer size in all NN correctors ($\rho^\mathrm{NN}_\mathrm{corr}(\mathcal{M})$, $\bar{\rho}^\mathrm{NN}_\mathrm{corr}(\Pi)$, $\bar{\rho}^\mathrm{NN}_\mathrm{M^2\mhyphen{}corr}(\mathcal{M})$) depends on $M$.
Due to its universality~\cite{Hornik1989,Grohs2023},
as our baseline model we choose a 6-layer fully connected (FC) network (multilayer perceptron) with 64 units (neurons) in each hidden layer and ReLU activations. The input/output layers are then adapted to meet the size $M$. 
Given the collection $\mathcal{M}$ of measurements of size $M$ and the training set of random density matrices $S_\mathrm{train}$ the NNs were trained to minimize the supervised training loss function:
\begin{equation}\label{eq:loss}
    \mathcal{L}(\mathcal{M},S_\mathrm{train}) = \frac{1}{N_b}\sum_{i}\Vert\rho_{i} - \rho^\mathrm{NN}_{i}(\mathcal{M})\Vert,
\end{equation}
where the difference between the original $\rho_i\in{}S_\mathrm{train}$ and the reconstructed $\rho^\mathrm{NN}$ is calculated averaging over a batch of size $N_b=64$, and $\Vert\cdot\Vert$ denotes a standard $L^2$ norm. The batch is a random fraction of $S_\mathrm{train}$ used in a single step of the training procedure (typically when training NN the stochastic gradient descent method~\cite{Grohs2023} is used). Random-sampling methods used to generate $S_\mathrm{train}$ and $S_\mathrm{test}$ sets are described in Appendix~B.

\subsection{Measurements selector network details}

\bluemark{The selector model consists of two recurrent neural networks. A recurrent network processes a sequence step by step and stores information from previous steps in an internal hidden state. Specifically, we utilize a popular LSTM~\cite{10.1162/neco_a_01199} architecture. Here, LSTM$_S$ is responsible for measurement selection, while LSTM$_R$ is responsible for state reconstruction. The input size of both networks is independent of the final number of measurements $M$, because the same cells are reused at each measurement step.}

Randomly selecting a reduced set of measurements is not the best possible approach. Having partial information about the system (i.e. few already performed measurements) may allow the model to predict the next most promising measurement operator, which gives the most valuable information about the system. Especially, we propose to feed the LSTM$_\mathrm{S}$ with initial measurement basis $\Pi_1={\mu_0}^{\otimes N}$ and force it to predict the next measurement operator $\Pi_2$, which is used to determine value $m_2$ according to Eq.~\ref{eq:measurements} in the main text. Subsequently, this procedure is repeated until the final choice $\Pi_M$ is obtained. 

Since there is no predetermined correct order of measurements, the typical approach of supervised training cannot be applied in this case, \bluemark{i.e., LSTM$_S$ cannot be trained by simply comparing its output with a prescribed correct sequence. Instead, LSTM$_S$ and LSTM$_R$ are trained together. At each step, LSTM$_R$ produces a reconstruction $\rho_l^{\rm NN}$, and the loss compares this reconstruction with the true training state. The loss is then summed over all steps, so that measurement choices are rewarded when they lead to better subsequent reconstructions.} 
To enable free information flow during \textit{backpropagation}~\cite{Grohs2023} (NN training), the \bluemark{measurement expectation values} algebra is incorporated directly into the NN architecture, \bluemark{i.e., values $m_l=\mathrm{Tr}(\Pi_l\rho)$ are calculated explicitly during the training phase. At the test stage, however, the reconstruction network does not require access to the target state $\rho$: the internally computed value $m_l$ can simply be replaced by an externally supplied experimental estimate $\hat m_l=n_l/N_c$.} Finally, both networks are trained with the use of $L_2$ loss
\begin{equation}
\mathcal{L}(S_\mathrm{train})=\frac{1}{N_bN_l}\sum_{i,l}\Vert\rho_i-\rho^\mathrm{NN}_{i,l}\Vert,\quad
\rho_i\in{}S_\mathrm{train},
\end{equation}
where $N_b$ denotes the number of samples in the $S_\mathrm{train}$ and $N_l$ is the number of selection-reconstruction steps. 
%Such a strategy enforces the network to return decent reconstructions even at early steps, hence increasing the chances of choosing the most important measurements first.

Two versions of the \textit{selector} network LSTM$_\mathrm{S}$ were proposed and tested. The first approach at step $l$ simply selects a previously unused measurement operator $\Pi_l$ from the catalog ${\cal C}_N$, defined in Eq.~(3) of the main text. The second strategy is to allow LSTM$_\mathrm{S}$ to produce a \textit{custom} measurement operator $\Pi_l$. The selected operator is then used to collect measurements and both the operator and the resulting measurement expectation values are passed to the reconstruction network LSTM$_\mathrm{R}$.
One should also note that the first version needs discrete selections that typically utilize the \textit{argmax} function, which is not differentiable. Therefore, to make the selection process differentiable and thus enable backpropagation, we applied the softmax function, i.e., the so-called \textit{Gumbel-softmax} activation~\cite{jang2016categorical}, to select the operator $\Pi_l$. However, further experiments have shown that another approach with an additional loss term for the \textit{selector} LSTM$_\mathrm{S}$, as discussed in Appendix~F, may be equally effective.
%The outputs of two different LSTM networks are then projected to either form of measurement operator in case of selector LSTM$_\mathrm{S}$ network and density matrix in case of reconstruction LSTM$_\mathrm{R}$ network.

We also note that, since the density matrix is complex, each NN processes the real and imaginary parts separately, treating them as distinct channels.

\section{Training and test datasets}
We generate a set of density matrices by using different random-sampling methods. 
The main idea is to generate random states using the quantum circuits approach, but in order to have a well-diversified ensemble, we also use a technique based on sampling from the uniform Haar measure \cite{mezzadri2006}. Additionally, to increase the number of maximally entangled states, we specifically include $20 000$ of them. 
All methods are described in detail in Ref.~\cite{Pawlowski2024}.
In total, we generate $460 000$ matrices for the training dataset $S_\mathrm{train}$ and $46 000$ matrices for the test set $S_\mathrm{test}$ for 2-qubit experiments. The sizes of the 3- and 4-qubit datasets were the same.
The datasets generated and/or analyzed during the current study are available in the Zenodo repository, \url{https://doi.org/10.5281/zenodo.17276582}.

\section{Single-qubit tomography from incomplete measurement set -- best approximation}
Here we analyze the simplest single-qubit system ($N=1$) and show how the standard quantum tomography scheme extended via the pseudoinverse operation into an incomplete measurements scenario, i.e. $M=2 < 4$, behaves and in what way this behavior can be corrected.

Let us assume a density matrix in the following form:
\begin{equation}
\rho=
  \begin{pmatrix}
    a & b\\
    b^\ast & 1\!-\!a
  \end{pmatrix},
  \label{eq:singlequbitrho}
\end{equation}
with real $a\in[0,1]$, complex $b$ with $|b|\in[0,\frac{1}{2}]$ and $a\geq a^2+|b|^2$ to fulfill positive semidefiniteness condition.
Then, the measurement expectation values (in the projector basis $\Pi_\nu=\mu_{i_1}=|\psi_{i_1}\rangle\langle \psi_{i_1}|$, with $|\psi_{i_1}\rangle$ (defined in Eq.~\ref{eq:single_qubit_catalogue}) are $(m_\nu)_{\nu=1}^4=(a,1-a,\frac{1}{2}+\mathrm{Re}(b),\frac{1}{2}+\mathrm{Im}(b))$, thus $B_{\nu\mu}$  matrix and its inverse are:
\begin{equation}
B_{\nu\mu}=
  \begin{pmatrix}
    1 & 0 & 0 & 1\\
    1 & 0 & 0 & -1\\
    1 & 1 & 0 & 0\\
    1 & 0 & -1 & 0\\
  \end{pmatrix}\!,\,
B_{\mu\nu}^{-1}=\frac{1}{2}
  \begin{pmatrix}
    1 & 1 & 0 & 0\\
    -1 & -1 & 2 & 0\\
    1 & 1 & 0 & -2\\
    1 & -1 & 0 & 0\\
  \end{pmatrix}\!.
  \label{eq:fullBsinglequbitexample}
\end{equation}
Finally, calculating $\rho=\Gamma_\mu B_{\mu\nu}^{-1}m_\nu$ restores the original matrix $\rho$ defined in Eq.~(\ref{eq:singlequbitrho}).

The situation changes if we reduce the number of projectors to two, e.g., $\nu=1,2$ by taking $\mu_1,\mu_2$ with $m_1=a$ and $m_2=1-a$ expectations. Now $B_{\nu\mu}$ [formed by two first rows of $B_{\nu\mu}$ from Eq.~(\ref{eq:fullBsinglequbitexample})] is no longer a square matrix and its inverse does not exist. However, one can calculate its pseudoinverse getting:
\begin{equation}
B_{\mu\nu}^+=\frac{1}{2}
  \begin{pmatrix}
    1 & 0 & 0 & 1\\
    1 & 0 & 0 & -1\\
  \end{pmatrix}^T\!,
\end{equation}
and then reconstruct the state via $\rho\simeq\Gamma_\mu B_{\mu\nu}^+m_\nu$ (note that $\nu=1,2$, $\mu=1,2,3,4$) formula:
\begin{equation}
\rho_\mathrm{pseudoinv}=
  \begin{pmatrix}
    a & 0\\
    0 & 1\!-\!a
  \end{pmatrix}\!.
  \label{eq:singlequbitrho12reconstructed}
\end{equation}
One can notice that the diagonal part of $\rho_\mathrm{pseudoinv}$ in Eq.~(\ref{eq:singlequbitrho12reconstructed}) is reconstructed correctly, while off-diagonal terms are absent. The most important observation is that since $m_1$ and $m_2$ measurement expectaion values contain only the information about populations, i.e. $a$, better reconstruction than Eq.~(\ref{eq:singlequbitrho12reconstructed}) is not possible and pseudoinverse scheme gives the best possible reconstruction and no corrections are needed.

However, when we select two other measurement expectation values: $\nu=1,3$, that is $m_1=a$ and $m_3=\frac{1}{2}+\mathrm{Re}(b)$, pseudoinverse will not give the best possible reconstruction.
Now the pseudoinverse is 
\begin{equation}
B_{\mu\nu}^+=\frac{1}{3}
  \begin{pmatrix}
    1 & -1 & 0 & 2\\
    1 & 2 & 0 & -1\\
  \end{pmatrix}^T\!,
\end{equation}
leading to the reconstruction:
\begin{equation}
\rho_\mathrm{pseudoinv}=\frac{1}{3}
  \begin{pmatrix}
    3a & 1-a+2\mathrm{Re}(b)\\
    1-a+2\mathrm{Re}(b) & 1-a+2\mathrm{Re}(b)\\
  \end{pmatrix}\!.
  \label{eq:singlequbitrho13reconstructed}
\end{equation}
that is not even correctly normalized. Nevertheless, if we add the following correction terms:
\begin{equation}
b_{\mu\nu}=\frac{1}{3}
  \begin{pmatrix}
    -1 & 1 & 0 & 1\\
    -1 & 1 & 0 & 1\\
  \end{pmatrix}^T\!,\,
c_\mu=\frac{1}{2}
  \begin{pmatrix}
    1 & -1 & 0 & -1
  \end{pmatrix}^T\!,
  \label{eq:13correctionterms}
\end{equation}
this via the formula $\rho\simeq\Gamma_\mu ((B_{\mu\nu}^+ + b_{\mu\nu})m_\nu+c_\mu)$ gives the best possible approximation:
\begin{equation}
\rho_\mathrm{corrected}=
  \begin{pmatrix}
    a & \mathrm{Re}(b)\\
    \mathrm{Re}(b) & 1-a\\
  \end{pmatrix}\!.
  \label{eq:singlequbitrho13corrected}
\end{equation}
This shows that there is room for action for the NN \textit{corrector} model compared to the scheme based on pseudoinverse. One can also verify that for terms defined in Eq.~(\ref{eq:13correctionterms}) vector $b_{\mu\nu}m_\nu+c_\mu=\frac{1-a-\mathrm{Re}(b)}{3}(1,-1,0,-1)^T$ is orthogonal to 1st and 3rd row in matrix $B_{\nu\mu}$ from Eq.~(\ref{eq:fullBsinglequbitexample}).
The same can be done for the other pairs of measurements obtaining various $\rho_\mathrm{corrected}$ for pairs of measurements $(\nu$, $\nu')$
\begin{gather}
\nu'\nonumber\\
\nu\,
  \begin{bmatrix}
  \begin{pmatrix}
    a & 0\\
    0 & 1-a
  \end{pmatrix}&
  \begin{pmatrix}
    a & \mathrm{Re}(b)\\
    \mathrm{Re}(b) & 1-a
  \end{pmatrix}&
  \begin{pmatrix}
    a & \mathrm{Im}(b)\\
    -\mathrm{Im}(b) & 1-a
  \end{pmatrix}\\ &
  \begin{pmatrix}
    a & \mathrm{Re}(b)\\
    \mathrm{Re}(b) & 1-a
  \end{pmatrix}&
  \begin{pmatrix}
    a & \mathrm{Im}(b)\\
    -\mathrm{Im}(b) & 1-a
  \end{pmatrix}\\ & &
  \begin{pmatrix}
    \frac{1}{2} & b\\
    b^\ast & \frac{1}{2}
  \end{pmatrix}  
  \end{bmatrix}\!,
  \label{eq:singlequbitbestanalytic}
\end{gather}
where rows denotes subsequent $\nu=1,2,3$ measurement expectations, while columns $\nu'=2,3,4$ expectation.
The expressions in Eq.~(\ref{eq:singlequbitbestanalytic}) were used to generate the results shown in Fig.~\ref{fig:1_qbit}(c) of the main text.

\section{2-qubit tomography -- optimal ordering for X states.}
The geometry of 2-qubit states is much more complicated than single-qubit ones~\cite{Omar2016,Jakubczyk2001,Kus2001}.
To verify effectiveness of NN reconstruction in 2-qubit case we introduce a family of so-called X states that generalize many important classes of mixed quantum states~\cite{Quesada2012}, but have much simpler structure than the general matrix:
\begin{equation}
X=\begin{pmatrix}
    a & 0 & 0 & w\\
    0 & b & z & 0\\
    0 & z^\ast & c & 0\\
    w^\ast & 0 & 0 & d
  \end{pmatrix}\!,
\end{equation}
with $a+b+c+d=1$, $a,b,c,d\geq 0$, and $|z|\leq\sqrt{bc}$, $|w|\leq\sqrt{ad}$.
Then the measurements in our basis, $\Pi_\nu=\mu_{i_1}\otimes\mu_{i_2}$, with $\mu_{i_k}=|\psi_{i_k}\rangle\langle \psi_{i_k}|$ defined in Eq.~\ref{eq:projector_catalogue}, are:
\begin{align}
(m_\nu&)_{\nu=1}^{16}=\\\nonumber
\big(&a,b,\frac{1}{2}(a+b), 
\frac{1}{2}(a+b),\\\nonumber
&c,d,\frac{1}{2}(c+d),
\frac{1}{2}(c+d),\\\nonumber
&\frac{1}{2}(a+c),
\frac{1}{2}(b+d),\\\nonumber
&\frac{1}{4}+\frac{1}{2}\left(\mathrm{Re}(w)+\mathrm{Re}(z)\right),
\frac{1}{4}+\frac{1}{2}\left((\mathrm{Im}(w)-\mathrm{Im}(z)\right),\\\nonumber
&\frac{1}{2}(a+c),
\frac{1}{2}(b+d),\\\nonumber
&\frac{1}{4}+\frac{1}{2}\left((\mathrm{Im}(w)+\mathrm{Im}(z)\right),\frac{1}{4}+\frac{1}{2}\left((\mathrm{Re}(z)-\mathrm{Re}(w)\right)\big).
\end{align}
Thus, some of the measurements are redundant, meaning that some orderings of measurements might be more optimal than others. This leaves room for the NN LSTM model to develop an optimization of the measurement sequence.

By training the LSTM on X states, we obtained sequences like $\nu=(1,2,6,11,16,12,15,...)$
or $(1,3,14,16,12,15,11,...)$. Typically, in the acquired sequences, the first three gave full information about the diagonal elements; while the next four (any permutation of $(11,12,15,16)$ measurements) gave full information about the coherences.
This result shows that there is a tomographic equivalent of the \textit{no free lunch} theorem~\cite{Wolpert1996,Wolpert1997}: by restricting the state space, we gain the possibility of more efficient QST by learning the optimal (redundancy-free) sequence of measurements. 

\section{Relation between correction terms}
\bluemark{
We start from Eq.~\ref{eq:corrector} by taking the matrix element
$\langle\psi_{\nu'}|\,\cdot\,|\psi_{\nu'}\rangle$
of both sides.
Assuming perfect reconstruction,
$\langle\psi_{\nu'}|\rho_{\mathrm{NN}}|\psi_{\nu'}\rangle \simeq m_{\nu'}$,
we obtain
\begin{equation}
m_{\nu'} =
B_{\nu'\mu}
\left[
\left(B^{+}_{\mu\nu}+b^{\mathrm{NN}}_{\mu\nu}\right)m_{\nu}
+ c^{\mathrm{NN}}_{\mu}
\right].
\end{equation}
This yields ($B_{\nu'\mu}B^{+}_{\mu\nu}=\delta_{\nu'\nu}$), 
\begin{equation}
m_{\nu'}=
\delta_{\nu'\nu}m_{\nu}
+B_{\nu'\mu}
\left(
b^{\mathrm{NN}}_{\mu\nu}m_{\nu}
+c^{\mathrm{NN}}_{\mu}
\right),
\end{equation}
and therefore
\begin{equation}
B_{\nu'\mu} v_{\mu}=0,
\end{equation}
where
\begin{equation}
v_{\mu}=b^{\mathrm{NN}}_{\mu\nu}m_{\nu}+c^{\mathrm{NN}}_{\mu},
\end{equation}
for all $\nu'=1,\ldots,M$.
This implies that the two corrector network outputs are not independent: the vector
$v_{\mu}$ must belong to the null space (kernel) of $B_{\nu'\mu}$, which is the orthogonal
complement of the row space of $B_{\nu'\mu}$.
}

\section{Loss in selector NN with predefined measurement basis}
The issue with the LSTM$_S$ model, which selects measurement operators from the predefined catalog ${\cal C}_N$ is that the \textit{argmax} function typically used in classifiers (selectors) is not differentiable.

\bluemark{The main approach that we utilize in this work is to use the softmax activation (e.g., Gumbel-softmax~\cite{jang2016categorical}) to construct the operator $\Pi_l'$ used only as input to LSTM$_\mathrm{R}$ during the training stage as:
\begin{equation}
    \Pi_l' = \sum_\nu \mathrm{Softmax}(P^l_\nu) \Pi_\nu,
\end{equation}
where $P^l$ is the probability vector returned by the LSTM$_\mathrm{S}$ at step $l$, containing the probabilities $P^l_\nu$ of choosing the measurement operator $\Pi_\nu$ from the discrete catalog ${\cal C}_N$. Additionally, we enforce the uniqueness of the choice, so that $\Pi_l'$ is different from previously selected operators $\{\Pi_1',...,\Pi_{l-1}'\}$. Importantly, while $\Pi_l'$ is used as LSTM$_R$ input, we keep $\Pi_l=\Pi_{\mathrm{argmax}(P^l)}$ as input of LSTM$_S$ and use it to determine the expectation value $m_l$. As a result, we allowed smooth gradient backpropagation through $\Pi_l'$ passed from LSTM$_S$ to LSTM$_R$.}

Simultaneously, a different approach resulted in a similar reconstruction accuracy. Here, during the training stage, in the given step $l$ each possible measurement pair ($\Pi_{\nu_l}$, $m_{\nu_l}$) is constructed and independently passed through the LSTM$_R$, resulting in set of reconstructed matrices $\{\rho^\mathrm{NN}_{\nu_l}\}_{\nu_l=1,\dots,4^N}$. Then,
$\mathcal{L}_{\nu_l}(S_\mathrm{train}) =\frac{1}{N_b}\sum_{i}\Vert\rho_i-\rho^\mathrm{NN}_{i,\nu_l}\Vert$, $\rho_i\in{}S_\mathrm{train}$
is calculated for each $\nu_l$ to find $\nu^*_l$ that minimizes the loss. Such $\nu^*_l$ is then set as a target in the cross-entropy loss for the vector $P^l$ returned by the LSTM$_S$:
\begin{equation}
    \mathcal{L}_\mathrm{CE}(P^l, \nu^*_l) = -\sum_{\nu_l} y_{\nu_l} \log(p_{\nu_l}),
\end{equation}
where $y_{\nu_l} = 1$ when $\nu_l$ is equal to $\nu^*_l$, and $y_{\nu_l} = 0$ otherwise.
Finally, the $\Pi_l$ is set as $\Pi_{\mathrm{argmax}(P^l)}$, meaning that the measurement operator $\Pi_l$ is chosen as the $\Pi_{\nu}$ with the highest probability $P^l_\nu$ in a given step $l$, but with additional restriction to preserve the uniqueness of measurements among all steps.  Such $\Pi_l$, together with respective $m_l$, are subsequently used as input for the reconstruction network LSTM$_\mathrm{R}$ trained using the $\mathcal{L}_l(S_\mathrm{train})$.

In contrast, the LSTM$_\mathrm{S}$ network that directly produces (a custom) measurement operator that can be passed to the reconstruction network, allows for straightforward simultaneous training of both LSTM$_\mathrm{S}$ and LSTM$_\mathrm{R}$ models using the single loss $\mathcal{L}(S_\mathrm{train})$. This approach allows for much faster training.
%and may lead to even greater exploitation of valuable measurements, on the other hand increasing the risk of producing hardly applicable measurement operators.

\bibliography{main}

@article{10.1162/neco_a_01199,
    author = {Yu, Yong and Si, Xiaosheng and Hu, Changhua and Zhang, Jianxun},
    title = "{A Review of Recurrent Neural Networks: LSTM Cells and Network Architectures}",
    journal = {Neural Computation},
    volume = {31},
    number = {7},
    pages = {1235-1270},
    year = {2019},
    month = {07},
    issn = {0899-7667},
    doi = {10.1162/neco_a_01199},
    url = {https://doi.org/10.1162/neco\_a\_01199},
    eprint = {https://direct.mit.edu/neco/article-pdf/31/7/1235/1053200/neco\_a\_01199.pdf},
}

@article{Kwiat2001,
  title = {Measurement of qubits},
  author = {James, Daniel F. V. and Kwiat, Paul G. and Munro, William J. and White, Andrew G.},
  journal = {Phys. Rev. A},
  volume = {64},
  issue = {5},
  pages = {052312},
  numpages = {15},
  year = {2001},
  month = {Oct},
  publisher = {American Physical Society},
  doi = {10.1103/PhysRevA.64.052312},
  url = {https://link.aps.org/doi/10.1103/PhysRevA.64.052312}
}

@book{Nielsen_Chuang_2010, 
place={Cambridge}, title={Quantum Computation and Quantum Information: 10th Anniversary Edition}, publisher={Cambridge University Press}, author={Nielsen, Michael A. and Chuang, Isaac L.}, year={2010}}

@article{Hradil2013,
  title={Informationally incomplete quantum tomography},
  author={Teo, Yong Siah and {\v{R}}eh{\'a}{\v{c}}ek, Jaroslav and Hradil, Zden{\u{e}}k},
  journal={Quantum Measurements and Quantum Metrology},
  volume={1},
  number={1},
  pages={57--83},
  year={2013},
  publisher={Versita}
}

@misc{yuan2022,
      title={A Brief Introduction to {POVM} Measurement in Quantum Communications}, 
      author={Renzhi Yuan},
      year={2022},
      eprint={2201.07968},
      archivePrefix={arXiv},
      primaryClass={quant-ph},
      url={https://arxiv.org/abs/2201.07968}, 
}

@article{Quesada2012,
author = {Nicolás Quesada, Asma Al-Qasimi and Daniel F.V. James},
title = {Quantum properties and dynamics of {X} states},
journal = {Journal of Modern Optics},
volume = {59},
number = {15},
pages = {1322--1329},
year = {2012},
publisher = {Taylor \& Francis},
doi = {10.1080/09500340.2012.713130},


URL = { 
    
        https://doi.org/10.1080/09500340.2012.713130
    
    

},
eprint = { 
    
        https://doi.org/10.1080/09500340.2012.713130
    
    

}
}

@article{Omar2016,
  title = {Entangled Bloch spheres: Bloch matrix and two-qubit state space},
  author = {Gamel, Omar},
  journal = {Phys. Rev. A},
  volume = {93},
  issue = {6},
  pages = {062320},
  numpages = {18},
  year = {2016},
  month = {Jun},
  publisher = {American Physical Society},
  doi = {10.1103/PhysRevA.93.062320},
  url = {https://link.aps.org/doi/10.1103/PhysRevA.93.062320}
}

@article{Jakubczyk2001,
title = {Geometry of Bloch vectors in two-qubit system},
journal = {Physics Letters A},
volume = {286},
number = {6},
pages = {383-390},
year = {2001},
issn = {0375-9601},
doi = {https://doi.org/10.1016/S0375-9601(01)00455-8},
url = {https://www.sciencedirect.com/science/article/pii/S0375960101004558},
author = {L. Jakóbczyk and M. Siennicki}
}

@article{Kus2001,
  title = {Geometry of entangled states},
  author = {Ku\ifmmode \acute{s}\else \'{s}\fi{}, Marek and \ifmmode \dot{Z}\else \.{Z}\fi{}yczkowski, Karol},
  journal = {Phys. Rev. A},
  volume = {63},
  issue = {3},
  pages = {032307},
  numpages = {13},
  year = {2001},
  month = {Feb},
  publisher = {American Physical Society},
  doi = {10.1103/PhysRevA.63.032307},
  url = {https://link.aps.org/doi/10.1103/PhysRevA.63.032307}
}

@article{Gilchrist2005,
  title = {Distance measures to compare real and ideal quantum processes},
  author = {Gilchrist, Alexei and Langford, Nathan K. and Nielsen, Michael A.},
  journal = {Phys. Rev. A},
  volume = {71},
  issue = {6},
  pages = {062310},
  numpages = {14},
  year = {2005},
  month = {Jun},
  publisher = {American Physical Society},
  doi = {10.1103/PhysRevA.71.062310},
  url = {https://link.aps.org/doi/10.1103/PhysRevA.71.062310}
}

@book{bengtsson2017,
	address = {Cambridge},
	author = {Bengtsson, Ingemar and Zyczkowski, Karol},
	db = {Cambridge Core},
	doi = {DOI: 10.1017/CBO9780511535048},
	dp = {Cambridge University Press},
	publisher = {Cambridge University Press},
	title = {Geometry of Quantum States: An Introduction to Quantum Entanglement},
	url = {https://www.cambridge.org/core/product/4BA9DCEED5BB16B222A917EAAAD17028},
	year = {2006}}

@article{jang2016categorical,
  title={Categorical reparameterization with gumbel-softmax},
  author={Jang, Eric and Gu, Shixiang and Poole, Ben},
  journal={arXiv preprint arXiv:1611.01144},
  year={2016}
}

@article{mezzadri2006,
author = {Mezzadri, Francesco},
year = {2006},
month = {10},
pages = {},
title = {How to generate random matrices from the classical compact groups},
volume = {54},
journal = {Notices of the American Mathematical Society}
}

@article{Pawlowski2024,
  title = {Identification of quantum entanglement with {Siamese} convolutional neural networks and semisupervised learning},
  author = {Paw\l{}owski, Jaros\l{}aw and Krawczyk, Mateusz},
  journal = {Phys. Rev. Appl.},
  volume = {22},
  issue = {1},
  pages = {014068},
  numpages = {13},
  year = {2024},
  month = {Jul},
  publisher = {American Physical Society},
  doi = {10.1103/PhysRevApplied.22.014068},
  url = {https://link.aps.org/doi/10.1103/PhysRevApplied.22.014068}
}

@article{Krawczyk2024,
  title = {Data-driven criteria for quantum correlations},
  author = {Krawczyk, Mateusz and Paw\l{}owski, Jaros\l{}aw and Ma\ifmmode \acute{s}\else \'{s}\fi{}ka, Maciej M. and Roszak, Katarzyna},
  journal = {Phys. Rev. A},
  volume = {109},
  issue = {2},
  pages = {022405},
  numpages = {12},
  year = {2024},
  month = {Feb},
  publisher = {American Physical Society},
  doi = {10.1103/PhysRevA.109.022405},
  url = {https://link.aps.org/doi/10.1103/PhysRevA.109.022405}
}

@article{Chen2022,
doi = {10.1088/2058-9565/ac310f},
url = {https://dx.doi.org/10.1088/2058-9565/ac310f},
year = {2021},
month = {nov},
publisher = {IOP Publishing},
volume = {7},
number = {1},
pages = {015005},
author = {Yiwei Chen and Yu Pan and Guofeng Zhang and Shuming Cheng},
title = {Detecting quantum entanglement with unsupervised learning},
journal = {Quantum Science and Technology}
}

@article{Asif2023,
	author = {Asif, Naema and Khalid, Uman and Khan, Awais and Duong, Trung Q. and Shin, Hyundong},
	date = {2023/01/28},
	doi = {10.1038/s41598-023-28745-3},
	id = {Asif2023},
	isbn = {2045-2322},
	journal = {Scientific Reports},
	number = {1},
	pages = {1562},
	title = {Entanglement detection with artificial neural networks},
	url = {https://doi.org/10.1038/s41598-023-28745-3},
	volume = {13},
	year = {2023}}

@article{Urena2024,
	author = {Ure{\~n}a, Julio and Sojo, Antonio and Bermejo-Vega, Juani and Manzano, Daniel},
	date = {2024/08/05},
	doi = {10.1038/s41598-024-68213-0},
	id = {Ure{\~n}a2024},
	isbn = {2045-2322},
	journal = {Scientific Reports},
	number = {1},
	pages = {18109},
	title = {Entanglement detection with classical deep neural networks},
	url = {https://doi.org/10.1038/s41598-024-68213-0},
	volume = {14},
	year = {2024}}

@article{Taghadomi2024,
	author = {Taghadomi, N. and Mani, A. and Fahim, A. and Bakouei, A.},
	date = {2025/03/12},
	doi = {10.1007/s42484-025-00267-3},
	id = {Taghadomi2025},
	isbn = {2524-4914},
	journal = {Quantum Machine Intelligence},
	number = {1},
	pages = {40},
	title = {Effective detection of quantum discord by using convolutional neural networks},
	url = {https://doi.org/10.1007/s42484-025-00267-3},
	volume = {7},
	year = {2025}}

@misc{quantum-tomography,
  title={{\tt Quantum-Tomography}: A python library to help perform tomography on a quantum state},
  author={Scott Turro},
  year={2023},
  howpublished={\url{https://quantumtomo.web.illinois.edu/}},
  note = {Version 1.0.7.0}
}

@article{Hornik1989,
  title={Multilayer feedforward networks are universal approximators},
  author={Hornik, Kurt and Stinchcombe, Maxwell and White, Halbert},
  journal={Neural networks},
  volume={2},
  number={5},
  pages={359--366},
  year={1989},
  publisher={Elsevier}
}

@book{Grohs2023,
  title={Mathematical aspects of deep learning},
  author={Grohs, Philipp and Kutyniok, Gitta},
  year={2023},
  publisher={Cambridge University Press},
  place={Cambridge, UK}
}

@article{Wolpert1996,
  title={The lack of a priori distinctions between learning algorithms},
  author={Wolpert, David H},
  journal={Neural computation},
  volume={8},
  number={7},
  pages={1341--1390},
  year={1996},
  publisher={MIT Press One Rogers Street, Cambridge, MA 02142-1209, USA journals-info~…}
}

@article{Hradil2001,
author = {Hradil, Zdeněk and Řeháček, Jaroslav},
title = {Efficiency of Maximum-likelihood Reconstruction of Quantum States},
journal = {Fortschritte der Physik},
volume = {49},
number = {10-11},
pages = {1083-1088},
doi = {https://doi.org/10.1002/1521-3978(200110)49:10/11<1083::AID-PROP1083>3.0.CO;2-K},
year = {2001}
}

@article{Hradil1997,
  title = {Quantum-state estimation},
  author = {Hradil, Z.},
  journal = {Phys. Rev. A},
  volume = {55},
  issue = {3},
  pages = {R1561--R1564},
  numpages = {0},
  year = {1997},
  month = {Mar},
  publisher = {American Physical Society},
  doi = {10.1103/PhysRevA.55.R1561},
  url = {https://link.aps.org/doi/10.1103/PhysRevA.55.R1561}
}

@Inbook{Buzek2004,
author="Bu{\v{z}}ek, Vladim{\'i}r",
editor="Paris, Matteo
and {\v{R}}eh{\'a}{\v{c}}ek, Jaroslav",
title="6 Quantum Tomography from Incomplete Data via MaxEnt Principle",
bookTitle="Quantum State Estimation",
year="2004",
publisher="Springer Berlin Heidelberg",
address="Berlin, Heidelberg",
pages="189--234",
isbn="978-3-540-44481-7",
doi="10.1007/978-3-540-44481-7_6",
url="https://doi.org/10.1007/978-3-540-44481-7_6"
}

@incollection{Altepeter2005,
title = {Photonic State Tomography},
editor = {P.R. Berman and C.C. Lin},
series = {Advances In Atomic, Molecular, and Optical Physics},
publisher = {Academic Press},
volume = {52},
pages = {105-159},
year = {2005},
issn = {1049-250X},
doi = {https://doi.org/10.1016/S1049-250X(05)52003-2},
url = {https://www.sciencedirect.com/science/article/pii/S1049250X05520032},
author = {J.B. Altepeter and E.R. Jeffrey and P.G. Kwiat}
}

@article{Vogel1989,
  title = {Determination of quasiprobability distributions in terms of probability distributions for the rotated quadrature phase},
  author = {Vogel, K. and Risken, H.},
  journal = {Phys. Rev. A},
  volume = {40},
  issue = {5},
  pages = {2847--2849},
  numpages = {0},
  year = {1989},
  month = {Sep},
  publisher = {American Physical Society},
  doi = {10.1103/PhysRevA.40.2847},
  url = {https://link.aps.org/doi/10.1103/PhysRevA.40.2847}
}

@book{Paris2004,
title = {Quantum state estimation},
editor = {Paris, Matteo and Rehacek, Jaroslav},
series = {Lecture Notes in Physics},
  volume={649},
  year={2004},
  publisher={Springer Berlin Heidelberg},
  address ={Berlin, Heidelberg}
}

@article{Cramer2010,
	author = {Cramer, Marcus and Plenio, Martin B. and Flammia, Steven T. and Somma, Rolando and Gross, David and Bartlett, Stephen D. and Landon-Cardinal, Olivier and Poulin, David and Liu, Yi-Kai},
	date = {2010/12/21},
	doi = {10.1038/ncomms1147},
	id = {Cramer2010},
	isbn = {2041-1723},
	journal = {Nature Communications},
	number = {1},
	pages = {149},
	title = {Efficient quantum state tomography},
	url = {https://doi.org/10.1038/ncomms1147},
	volume = {1},
	year = {2010}}

@article{Acharya2021,
  title = {Shadow tomography based on informationally complete positive operator-valued measure},
  author = {Acharya, Atithi and Saha, Siddhartha and Sengupta, Anirvan M.},
  journal = {Phys. Rev. A},
  volume = {104},
  issue = {5},
  pages = {052418},
  numpages = {11},
  year = {2021},
  month = {Nov},
  publisher = {American Physical Society},
  doi = {10.1103/PhysRevA.104.052418},
  url = {https://link.aps.org/doi/10.1103/PhysRevA.104.052418}
}

@article{Huang2020,
	author = {Huang, Hsin-Yuan and Kueng, Richard and Preskill, John},
	date = {2020/10/01},
	doi = {10.1038/s41567-020-0932-7},
	id = {Huang2020},
	isbn = {1745-2481},
	journal = {Nature Physics},
	number = {10},
	pages = {1050--1057},
	title = {Predicting many properties of a quantum system from very few measurements},
	url = {https://doi.org/10.1038/s41567-020-0932-7},
	volume = {16},
	year = {2020}}

@article{Carrasquilla2019,
	author = {Carrasquilla, Juan and Torlai, Giacomo and Melko, Roger G. and Aolita, Leandro},
	date = {2019/03/01},
	doi = {10.1038/s42256-019-0028-1},
	id = {Carrasquilla2019},
	isbn = {2522-5839},
	journal = {Nature Machine Intelligence},
	number = {3},
	pages = {155--161},
	title = {Reconstructing quantum states with generative models},
	url = {https://doi.org/10.1038/s42256-019-0028-1},
	volume = {1},
	year = {2019}}

@article{Aaronson2020,
author = {Aaronson, Scott},
title = {Shadow Tomography of Quantum States},
journal = {SIAM Journal on Computing},
volume = {49},
number = {5},
pages = {STOC18-368-STOC18-394},
year = {2020},
doi = {10.1137/18M120275X},

URL = { 
    
        https://doi.org/10.1137/18M120275X
    
    

},
eprint = { 
    
        https://doi.org/10.1137/18M120275X
    
    

}
}

@article{Opatrny1997,
  title = {Least-squares inversion for density-matrix reconstruction},
  author = {Opatrn\'y, T. and Welsch, D.-G. and Vogel, W.},
  journal = {Phys. Rev. A},
  volume = {56},
  issue = {3},
  pages = {1788--1799},
  numpages = {0},
  year = {1997},
  month = {Sep},
  publisher = {American Physical Society},
  doi = {10.1103/PhysRevA.56.1788},
  url = {https://link.aps.org/doi/10.1103/PhysRevA.56.1788}
}

@article{Teo2011,
  title = {Quantum-State Reconstruction by Maximizing Likelihood and Entropy},
  author = {Teo, Yong Siah and Zhu, Huangjun and Englert, Berthold-Georg and \ifmmode \check{R}\else \v{R}\fi{}eh\'a\ifmmode \check{c}\else \v{c}\fi{}ek, Jaroslav and Hradil, Zdenek},
  journal = {Phys. Rev. Lett.},
  volume = {107},
  issue = {2},
  pages = {020404},
  numpages = {4},
  year = {2011},
  month = {Jul},
  publisher = {American Physical Society},
  doi = {10.1103/PhysRevLett.107.020404},
  url = {https://link.aps.org/doi/10.1103/PhysRevLett.107.020404}
}

@ARTICLE{Wolpert1997,
  author={Wolpert, D.H. and Macready, W.G.},
  journal={IEEE Transactions on Evolutionary Computation}, 
  title={No free lunch theorems for optimization}, 
  year={1997},
  volume={1},
  number={1},
  pages={67-82},
  doi={10.1109/4235.585893}}

@article{Torlai2018,
	author = {Torlai, Giacomo and Mazzola, Guglielmo and Carrasquilla, Juan and Troyer, Matthias and Melko, Roger and Carleo, Giuseppe},
	date = {2018/05/01},
	doi = {10.1038/s41567-018-0048-5},
	id = {Torlai2018},
	isbn = {1745-2481},
	journal = {Nature Physics},
	number = {5},
	pages = {447--450},
	title = {Neural-network quantum state tomography},
	url = {https://doi.org/10.1038/s41567-018-0048-5},
	volume = {14},
	year = {2018}}

@article{Kotuny2022,
  title = {Neural-network quantum state tomography},
  author = {Koutn\'y, Dominik and Motka, Libor and Hradil, Zden\ifmmode \check{e}\else \v{e}\fi{}k and \ifmmode \check{R}\else \v{R}\fi{}eh\'a\ifmmode \check{c}\else \v{c}\fi{}ek, Jaroslav and S\'anchez-Soto, Luis L.},
  journal = {Phys. Rev. A},
  volume = {106},
  issue = {1},
  pages = {012409},
  numpages = {8},
  year = {2022},
  month = {Jul},
  publisher = {American Physical Society},
  doi = {10.1103/PhysRevA.106.012409},
  url = {https://link.aps.org/doi/10.1103/PhysRevA.106.012409}
}

@article{Xin2019,
	author = {Xin, Tao and Lu, Sirui and Cao, Ningping and Anikeeva, Galit and Lu, Dawei and Li, Jun and Long, Guilu and Zeng, Bei},
	date = {2019/11/29},
	doi = {10.1038/s41534-019-0222-3},
	id = {Xin2019},
	isbn = {2056-6387},
	journal = {npj Quantum Information},
	number = {1},
	pages = {109},
	title = {Local-measurement-based quantum state tomography via neural networks},
	url = {https://doi.org/10.1038/s41534-019-0222-3},
	volume = {5},
	year = {2019}}

@article{Penrose1955, title={A generalized inverse for matrices}, volume={51}, DOI={10.1017/S0305004100030401}, number={3}, journal={Mathematical Proceedings of the Cambridge Philosophical Society}, author={Penrose, R.}, year={1955}, pages={406–413}}

@article{Li2017,
  title = {Optimal design of measurement settings for quantum-state-tomography experiments},
  author = {Li, Jun and Huang, Shilin and Luo, Zhihuang and Li, Keren and Lu, Dawei and Zeng, Bei},
  journal = {Phys. Rev. A},
  volume = {96},
  issue = {3},
  pages = {032307},
  numpages = {6},
  year = {2017},
  month = {Sep},
  publisher = {American Physical Society},
  doi = {10.1103/PhysRevA.96.032307},
  url = {https://link.aps.org/doi/10.1103/PhysRevA.96.032307}
}

@article{Shahnawaz2021,
  title = {Quantum State Tomography with Conditional Generative Adversarial Networks},
  author = {Ahmed, Shahnawaz and S\'anchez Mu\~noz, Carlos and Nori, Franco and Kockum, Anton Frisk},
  journal = {Phys. Rev. Lett.},
  volume = {127},
  issue = {14},
  pages = {140502},
  numpages = {8},
  year = {2021},
  month = {Sep},
  publisher = {American Physical Society},
  doi = {10.1103/PhysRevLett.127.140502},
  url = {https://link.aps.org/doi/10.1103/PhysRevLett.127.140502}
}

@article{Schmale2022,
	author = {Schmale, Tobias and Reh, Moritz and G{\"a}rttner, Martin},
	date = {2022/09/23},
	doi = {10.1038/s41534-022-00621-4},
	id = {Schmale2022},
	isbn = {2056-6387},
	journal = {npj Quantum Information},
	number = {1},
	pages = {115},
	title = {Efficient quantum state tomography with convolutional neural networks},
	url = {https://doi.org/10.1038/s41534-022-00621-4},
	volume = {8},
	year = {2022}}

@article{Quek2021,
	author = {Quek, Yihui and Fort, Stanislav and Ng, Hui Khoon},
	date = {2021/06/24},
	doi = {10.1038/s41534-021-00436-9},
	id = {Quek2021},
	isbn = {2056-6387},
	journal = {npj Quantum Information},
	number = {1},
	pages = {105},
	title = {Adaptive quantum state tomography with neural networks},
	url = {https://doi.org/10.1038/s41534-021-00436-9},
	volume = {7},
	year = {2021}}

@article{Melkani2020,
  title = {Eigenstate extraction with neural-network tomography},
  author = {Melkani, Abhijeet and Gneiting, Clemens and Nori, Franco},
  journal = {Phys. Rev. A},
  volume = {102},
  issue = {2},
  pages = {022412},
  numpages = {11},
  year = {2020},
  month = {Aug},
  publisher = {American Physical Society},
  doi = {10.1103/PhysRevA.102.022412},
  url = {https://link.aps.org/doi/10.1103/PhysRevA.102.022412}
}

@article{Palmieri2024,
  title = {Enhancing quantum state tomography via resource-efficient attention-based neural networks},
  author = {Palmieri, Adriano Macarone and M\"uller-Rigat, Guillem and Srivastava, Anubhav Kumar and Lewenstein, Maciej and Rajchel-Mieldzio\ifmmode \acute{c}\else \'{c}\fi{}, Grzegorz and P\l{}odzie\ifmmode \acute{n}\else \'{n}\fi{}, Marcin},
  journal = {Phys. Rev. Res.},
  volume = {6},
  issue = {3},
  pages = {033248},
  numpages = {16},
  year = {2024},
  month = {Sep},
  publisher = {American Physical Society},
  doi = {10.1103/PhysRevResearch.6.033248},
  url = {https://link.aps.org/doi/10.1103/PhysRevResearch.6.033248}
}

@article{Giza2010,
  title = {Permutationally Invariant Quantum Tomography},
  author = {T\'oth, G. and Wieczorek, W. and Gross, D. and Krischek, R. and Schwemmer, C. and Weinfurter, H.},
  journal = {Phys. Rev. Lett.},
  volume = {105},
  issue = {25},
  pages = {250403},
  numpages = {4},
  year = {2010},
  month = {Dec},
  publisher = {American Physical Society},
  doi = {10.1103/PhysRevLett.105.250403},
  url = {https://link.aps.org/doi/10.1103/PhysRevLett.105.250403}
}

@article{Giza2014,
  title = {Experimental Comparison of Efficient Tomography Schemes for a Six-Qubit State},
  author = {Schwemmer, Christian and T\'oth, G\'eza and Niggebaum, Alexander and Moroder, Tobias and Gross, David and G\"uhne, Otfried and Weinfurter, Harald},
  journal = {Phys. Rev. Lett.},
  volume = {113},
  issue = {4},
  pages = {040503},
  numpages = {5},
  year = {2014},
  month = {Jul},
  publisher = {American Physical Society},
  doi = {10.1103/PhysRevLett.113.040503},
  url = {https://link.aps.org/doi/10.1103/PhysRevLett.113.040503}
}

@article{Gross2010,
  title = {Quantum State Tomography via Compressed Sensing},
  author = {Gross, David and Liu, Yi-Kai and Flammia, Steven T. and Becker, Stephen and Eisert, Jens},
  journal = {Phys. Rev. Lett.},
  volume = {105},
  issue = {15},
  pages = {150401},
  numpages = {4},
  year = {2010},
  month = {Oct},
  publisher = {American Physical Society},
  doi = {10.1103/PhysRevLett.105.150401},
  url = {https://link.aps.org/doi/10.1103/PhysRevLett.105.150401}
}

@article{Balaz2025,
  title = {Learning entanglement from tomography data: Contradictory measurement importance for neural networks and random forests},
  author = {Bal\'a\ifmmode \check{z}\else \v{z}\fi{}, Pavel and Krawczyk, Mateusz and Paw\l{}owski, Jaros\l{}aw and Roszak, Katarzyna},
  journal = {Phys. Rev. A},
  volume = {112},
  issue = {2},
  pages = {022431},
  numpages = {13},
  year = {2025},
  month = {Aug},
  publisher = {American Physical Society},
  doi = {10.1103/zgfy-yyw8},
  url = {https://link.aps.org/doi/10.1103/zgfy-yyw8}
}

@article{Blume2010,
  title = {Hedged Maximum Likelihood Quantum State Estimation},
  author = {Blume-Kohout, Robin},
  journal = {Phys. Rev. Lett.},
  volume = {105},
  issue = {20},
  pages = {200504},
  numpages = {4},
  year = {2010},
  month = {Nov},
  publisher = {American Physical Society},
  doi = {10.1103/PhysRevLett.105.200504},
  url = {https://link.aps.org/doi/10.1103/PhysRevLett.105.200504}
}

\end{document}